\documentclass[nofootinbib,prx,amsmath,amssymb,onecolumn,floatfix]{revtex4-2}
\usepackage{graphicx}
\usepackage{dcolumn}
\usepackage{bm}
\usepackage{color}
\usepackage{siunitx}

\newcommand\blfootnote[1]{%
  \begingroup
  \renewcommand\thefootnote{}\footnote{#1}%
  \addtocounter{footnote}{-1}%
  \endgroup
}
\usepackage{lineno}

\begin{document}

\preprint{APS/123-QED}
\title{Observation of Elastic Orbital Angular Momentum Transfer: \\ Coupling Flexural Waves in Partially Submerged Pipes to Acoustic Waves in Fluids}

\author{G.~J. Chaplain$^{1*}$, J.~M. De Ponti$^2$ and T.~A. Starkey$^{1}$}
\affiliation{$^1$Electromagnetic and Acoustic Materials Group, Department of Physics and Astronomy, University of Exeter, Exeter EX4 4QL, United Kingdom}
\affiliation{$^2$Department of Civil and Environmental Engineering, Politecnico di Milano, Piazza Leonardo da Vinci, 32, 20133 Milano, Italy}
\date{\today}

\begin{abstract}
Research into the orbital angular momentum carried by helical wave-fronts has been dominated by the fields of electromagnetism and acoustics, owing to its practical utility in sensing, communication and tweezing. Despite the huge research effort across the wave community, only recently has \textit {elastic} orbital angular momentum been theoretically shown to exhibit similar properties. Here we experimentally observe the transfer of elastic orbital angular momentum from a hollow elastic pipe to a fluid in which the pipe is partially submerged, in an elastic analogue of Durnin’s slit-ring experiment for optical beams. This transfer is achieved by coupling the dilatational component of guided flexural waves in the pipe with the pressure field in the fluid; the circumferential distribution of the normal stress in the pipe acts as a continuous phased pressure source in the fluid resulting in the generation of Bessel-like acoustic beams. This demonstration has implications for future research into a new regime of orbital angular momentum for elastic waves, as well providing a new method to excite acoustic beams that carry orbital angular momentum that could create a new paradigm shift for acoustic tweezing.
\end{abstract}

\maketitle

\blfootnote{$^*$ Corresponding Author: {g.j.chaplain@exeter.ac.uk}}
Over the past three decades investigation into the orbital angular momentum (OAM) carried by helical waves has largely been reserved for light and sound \cite{shen2019optical}. The realisation by Allen \textit{et al.} \cite{allen1992orbital} that electromagnetic Laguerre-Gaussian (LG) beams, satisfying the paraxial wave equation \cite{fontaine2019laguerre}, carry a well defined OAM sparked a resurgence in interest in exploiting this mechanical property of light for optical tweezers \cite{yao2011orbital,loke2014driving,allen2016optical,padgett2017orbital,barnett2017optical}. The celebrated success of utilising LG beams comes, in part, from the ease in which they can be generated; a variety of simple devices can form these modes, for example spiral phase plates, q-plates, and spatial light modulators  \cite{LaserFocusWorld,oemrawsingh2004production}. The  orbital angular momentum is associated with the spatial distribution of the beam, and not with the polarisation (that determines the intrinsic spin angular momentum) \cite{allen1992orbital}; the inclined phase-fronts give rise to a well defined OAM about the beam axis that is proportional to the azimuthal index, or topological charge, $m$. The azimuthal variation of phase forms a helical profile varying as $e^{\pm im\theta}$, for azimuthal angle $\theta$ with the sign of $m$ determining the handedness of the helix. In general any optical beam with inclined phase fronts can carry a well-defined OAM \cite{o2002intrinsic}. Since this realisation other forms of optical beams that carry OAM have been investigated, by interference and superposition, such as higher-order Bessel beams \cite{mcgloin2003interfering,vasilyeu2009generating}, and more general vortex structures (knots) with non-integer OAM \cite{leach2004knotted,leach2005vortex,gotte2008light,morgan2017higher}. 

Translations of the phenomena surrounding OAM beams have been shown to hold for sound \cite{bliokh2019spin}, with the transfer of torque for scalar waves within the non-paraxial regime being considered \cite{zhang2011angular,zhang2018reversals}. Similar applications in acoustic tweezers exist for micro-scale biological manipulation and communication, to name a few \cite{hong2015observation,baresch2016observation,shi2017high,marzo2018acoustic}. Conventionally the excitation of acoustic OAM modes rests on acoustic analogues of phase plates \cite{hefner1998acoustical,hefner1999acoustical}, resonant devices and antenna concepts \cite{jiang2016convert,naify2016generation,guo2019high}, or by discrete phased arrays \cite{gibson2018reversal,antonacci2019demonstration,cromb2020amplification}. Utilising discrete phased arrays is indeed a favoured method for the excitation of higher-order OAM acoustic modes, that carry topological charge $m >1$ \cite{marchiano2005synthesis}. Importantly, such higher-order modes exhibit an instability phenomenon: a vortex of charge $m > 1$ is not stable and degenerates into $|m|$ vortices (screw dislocations) of charge $m/|m|$ \cite{basistiy1993optics,zambrini2006quasi,allen2016optical}.

Recently the consideration of OAM has been extended further into the realm of elasticity, where both the spin-phonon coupling in elastic media and the intrinsic spin of elastic waves have been investigated \cite{nakane2018angular,long2018intrinsic}. The orbital angular momentum associated with only the compressional potential of flexural waves in elastic pipes has also been considered \cite{Chaplain2022eOAM}, with the complete theory shown by Bliokh \cite{bliokh2022elastic}, proving that the \textit{total} angular momentum (the sum of spin and orbital contributions) is quantised, as in electromagnetism and acoustics. In this paper we experimentally verify the experiment proposed in \cite{Chaplain2022eOAM}, showing that the elastic OAM associated flexural waves can be coupled to a fluid, thereby observing its transfer through the excitation of acoustic pressure fields that carry OAM. Guided ultrasonic modes in pipes are considered, specifically flexural modes with a natural circumferential variation of phase. The coupling to the acoustic pressure in a fluid is achieved by partially submerging one end of the pipe in the fluid, in a set up resembling an elastic analogue of Durnin's slit-ring experiment \cite{durnin1987diffraction,vasilyeu2009generating}. We show that the generation of Bessel-like, higher-order acoustic OAM modes is possible, with clear agreement between simulation and experiments, highlighting specifically the features associated for an $m = 3$ topological charge. Before introducing the flexural modes and the pipe structure we recall the governing equations of elastic materials and the associated compressional OAM. 

An isotropic, homogeneous linear elastic material supports waves governed by the Navier-Cauchy equations \cite{landau1959course}, following the Einstein summation convention,
\begin{equation}
    \mu\partial_{j}\partial_{j}\xi_{i} + (\lambda + \mu)\partial_{j}\partial_{i}\xi_{i} = \rho\ddot{\xi_{i}},
    \label{eq:elastod}
\end{equation}
and the constitutive law
\begin{align}
\begin{split}
    \sigma_{ij} &= C_{ijkl}\varepsilon_{kl} = \lambda\delta_{ij}\varepsilon_{kk} + 2\mu\varepsilon_{ij},
    \end{split}
\end{align}
with $\xi_{i}$ the displacement and $\ddot{\xi_{i}}$ its double time derivative; Lam\'{e}'s first and second parameters are denoted $\lambda$, $\mu$ respectively with $\rho$ being the material density; $\sigma_{ij}$ and $C_{ijkl}$ are the stress and stiffness tensors respectively; and $\varepsilon_{ij} \equiv \frac{1}{2}(\xi_{i,j} + \xi_{j,i})$ is the strain tensor (comma notation denotes partial differentiation, and $\partial_i \equiv \frac{\partial}{\partial x_{i}}$).

The displacement comprises both shear and compressional motion, described by an equivoluminal vector potential $\Psi_{i}$ and scalar dilatational potential $\Phi$ respectively such that, by Helmholtz decomposition, $\xi_i = \partial_{i}\Phi + \epsilon_{ijk}\partial_{j}\Psi_{k}$. Elastic waves with inclined phase-fronts naturally occur as flexural modes in pipe walls \cite{Chaplain2022eOAM}. Choosing cylindrical coordinates oriented with $z$ along the pipe axis the potentials take the form
\begin{align}
    \begin{split}
     \Phi &= \phi(r)\exp \left[i(m\theta + k_{z}z - \omega t)\right], \\
     \Psi_{\alpha} &= \psi_{\alpha}(r) \exp \left[i(m\theta + k_{z}z - \omega t)\right],\\ 
     \label{eq:ansatz}
\end{split}
\end{align}
where $\alpha = r,\theta,z$. $k_z$ is the wave number along the pipe axis and $\omega$ the radian frequency. Stress free boundary conditions are imposed on the inner and outer radii $r_a$, $r_b$ respectively, such that $\sigma_{rr} = \sigma_{r\theta} = \sigma_{rz} \vert_{r_{a,b}} = 0$, along with the infinite cylinder gauge condition $\nabla\cdotp\boldsymbol{\Psi} = 0$. The resulting radial distribution of the dilatational and shear potentials ($\phi(r)$ and $\psi_{\alpha}(r)$) are then described by a linear combination of Bessel functions and their modifications \cite{gazis1959a}. We consider the coupling of elastic waves in a pipe with pressure fields in a fluid that the pipe is partially submerged. At the solid-fluid boundary the acoustic pressure induces a fluid load on the solid structure, and the structural acceleration acts as a normal acceleration across the solid-fluid boundary. The boundary conditions then manifest as the dynamic continuity of traction, and the kinematic continuity of the normal particle displacement at the interface between the solid and fluid, such that \cite{morand1995fluid}

\begin{align}
\begin{split}
 \sigma_{ij}n_{j} &= -pI_{ij}n_{j},\\
 \xi_in_{i} &= u_in_{i},
\end{split}
 \end{align}
where $p$ is the acoustic pressure, $I_{ij}$ is the unit tensor, $u_{i}$ is the particle displacement in the fluid and $n_j$ is the surface normal. In the fluid domain the governing equations follow the standard linearised Euler equations. For harmonic motion these read
\begin{align}
    \begin{split}
        \partial_{i}p - \omega^2\rho_{F}u_{i} &= 0, \\
        p + \rho_{F}c_{F}^{2}\partial_{i}u_{i} &= 0,
    \end{split}
\end{align}
for fluid density $\rho_{F}$ with the acoustic wavespeed in the fluid denoted $c_{F}$. The coupling boundary conditions of the elastoacoustic problem can then be expressed in the pure displacement formulation \cite{bermudez2008fluid}:
\begin{align}
    \begin{split}
        \sigma_{ij}n_{j} - \rho_{F}c_{F}^2\partial_{k}u_{k}I_{ij}n_{j} &=0, \\
        \xi_{i}n_{i} - u_{i}n_{i} &= 0.
    \end{split}
\end{align}
These equations are solved numerically throughout via the Finite Element Method (FEM) \cite{comsolSolidMech}.

The vector system of elasticity supports two body waves, compression and shear, that travel with distinct wavespeeds given by $c_p = \sqrt{(\lambda + 2\mu)/\rho}$ and $c_s = \sqrt{\mu/\rho}$ respectively. Clearly, in the fully coupled elastic system, longitudinal (compressional), transverse (shear) and hybrid components all contribute to the orbital angular momentum, and thus to the total angular momentum (orbital and spin) \cite{bliokh2022elastic}. Therefore, the splitting of the displacement vector into the dilatational and shear potentials, as done in \cite{Chaplain2022eOAM}, is not required; however, only considering the OAM associated with the compressional potential (that is proportional to the azimuthal index $m$) motivated this experimental work as fluids only support compressional waves and not shear.

Optical and acoustic tweezing are primary applications for the transfer of OAM, where particles can be trapped and manipulated at the vortex singularity at the beam centre \cite{franke2008advances,hong2015observation}. In the case of elastic OAM carried by flexural modes in pipes there is no such singularity as the pipe is hollow - a direct analogy cannot be drawn in this case as there is no elastic medium to suspend particles with at the pipe centre, along its axis. However, it is well known that vibrating solids radiate sound waves in fluids, and it is this coupling we leverage to excite an acoustic OAM mode via the flexural displacement field carried in the elastic pipe. We experimentally validate this OAM transfer, thereby developing a new continuous-phased acoustic source in the form of flexural modes in pipes; Bessel-like beams are generated following from the radial distribution of the compressional potential.

We first detail the background of flexural modes in pipes and how they are efficiently generated, using an elastic analogue to optical spiral phase plates. We utilise this analogue, the so-called elastic spiral phase pipe \cite{Chaplain2022eSPP}, to show the first experimental observation of the transfer of elastic OAM from flexural modes in pipes to fluids. We include qualitative comparisons with both Finite Element Method (FEM) simulations and Dynamic Mode Decomposition (DMD), and a comparison to classical discrete phased sources, before concluding and highlighting perspectives for applications. 

\section*{Guided Flexural Modes in Pipes} \label{sec:pipes}
Guided ultrasonic waves in pipes have long been studied, with the first analytical description being posed by Gazis in the late 1950s \cite{gazis1959a,gazis1959b}, for infinitely long pipes. They fall into three modal classes: longitudinal ($L$), torsional ($T$), and flexural ($F$), with a naming convention attributed to Silk and Bainton \cite{silk1979propagation} such that they are written $L(m,n)$, $T(m,n)$ and $F(m,n)$. The integers $m$, $n$ denote the circumferential and group order respectively; the circumferential order being analogous to the topological charge of optical vortex beams. The generation and inspection of these guided waves has found much success in non-destructive techniques and evaluation \cite{Alleyne98,Lowe98}. The focus of this paper is to observe the coupling between non-axisymmetric flexural modes $F(m > 0,n)$ in pipes and acoustic waves in a fluid. As such we require a device to ensure their efficient generation. 

The recent advent of the elastic spiral phase pipe (eSPP) \cite{Chaplain2022eSPP} achieves this. This structure removes the necessity to rely on conventional means of complex arrangements of transducers or phased arrays (e.g by comb arrays or non-axisymmetric partial loading \cite{shin1999guided,tang2017excitation}). Advantages of the eSPP include passively exciting arbitrary flexural modes with, crucially, single handedness (i.e. only one sign of $m$) by mode converting longitudinal modes (e.g. $L(0,2)$) that can be easily to excited in isolation \cite{LOWE20011551}. This is a particularly attractive property as a candidate for using flexural modes that are sensitive to axial cracks, where the conventional longitudinal and torsional modes are weakly sensitive \cite{kwun2008detection,ratassepp2010scattering}.

\begin{figure}
    \centering
    \includegraphics[width = 0.9\textwidth]{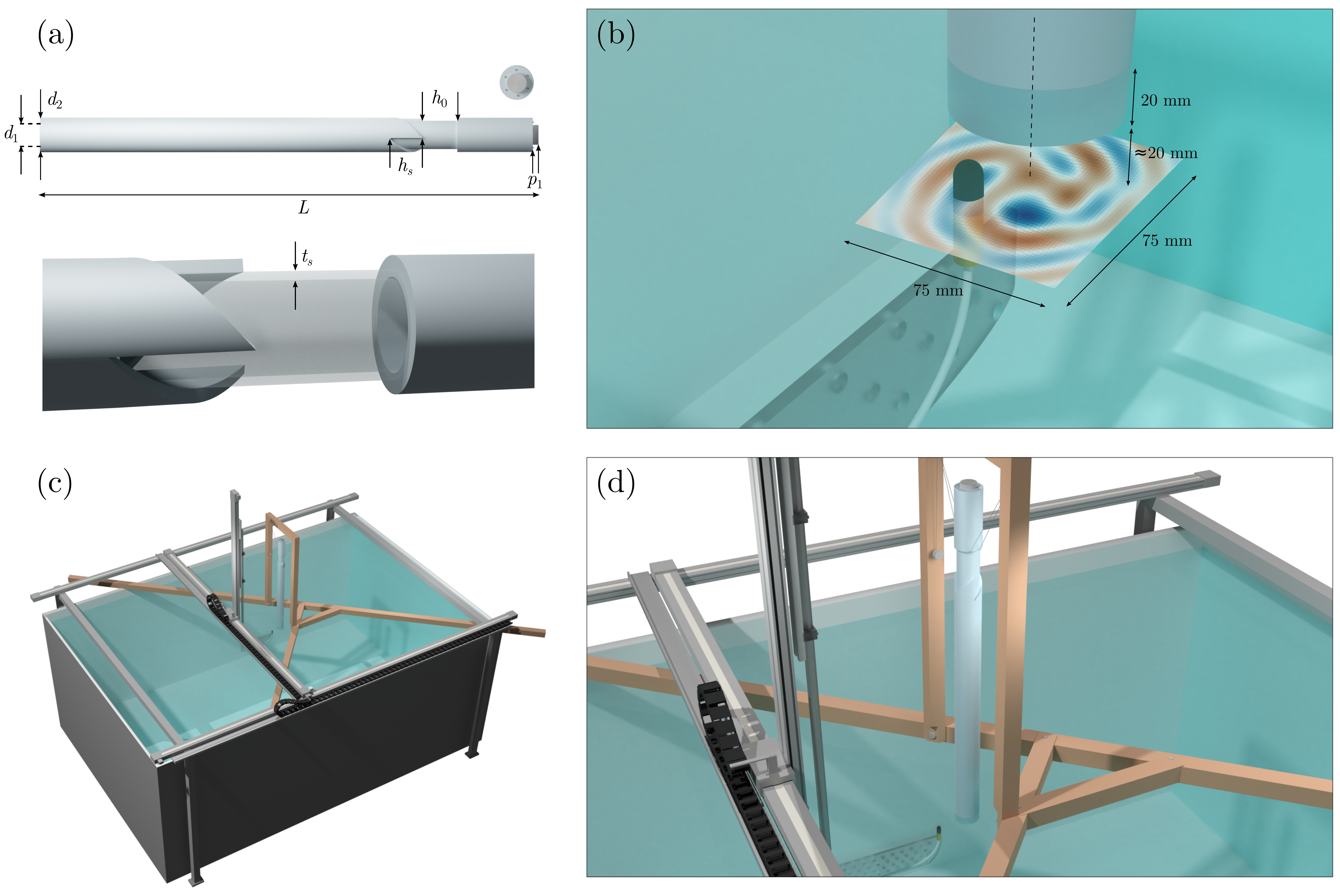}
    \caption{(a) Elastic Spiral Phase Pipe: $L = 900$ mm, $d_1 = 40$ mm, $d_2 = 60$ mm, $h_s = 63$ mm, $h_0 = 95$ mm, $p_1 = 12$ mm and the thickness at the spiral region, $t_s = 4$ mm (shown in zoom, where spiral region has been made partially transparent). Also shown is a View of the capped end of the pipe also shown with the piezo disk, of diameter $35$ mm, and attaching screws. (b) Acoustic Measurement Centre: Schematic of close-up of the submerged end of the pipe showing orientation of the hydrophone. The measurement plane is approximately $20~\si{\milli\meter}$ below the end of the pipe, intersecting the detector at its `acoustic centre'. Shown too is an example experimental profile after frequency domain analysis. (c) water tank and $xyz$ scanning stage (d) Close-up of suspended pipe and scanning arm.}
    \label{fig:Pipe}
\end{figure}

In Fig.~\ref{fig:Pipe} we detail the eSPP used in the experimental verification of the transfer of elastic OAM. In Fig.~\ref{fig:Pipe}(a) we show the elastic spiral phase pipe that endows incoming axisymmetric waves with a helical phase profile, similar to OAM generation by optical and acoustic analogues \cite{beijersbergen1994helical,hefner1998acoustical,hefner1999acoustical}. The spiral pipe used here has already been characterised \cite{Chaplain2022eSPP}, comprising an aluminium pipe of density $\rho = 2710~\si{\kilo\gram\meter^{-3}}$, Young's Modulus $E = 70~\si{\giga\pascal}$ and Poisson's ratio $\nu = 0.33$. The inner and outer diameters of the pipe are $d_1 = 40~\si{\milli\meter}$ and $d_2 = 60~\si{\milli\meter}$ respectively. One end of the pipe is open (to be submerged in fluid), while the other is capped with an aluminium disk of diameter $d_2$ and thickness $10~\si{\milli\meter}$, attached by six screws. The total length of the pipe is $L = 900~\si{\milli\meter}$. The spiral region of the pipe is specifically designed to convert $L(0,2)$ modes to $F(3,2)$ modes at $62~\si{\kilo\hertz}$, and is formed of by CNC milling a thickness of $6~\si{\milli\meter}$ from the pipe into three spiral steps of length $h_s = 63~\si{\milli\meter}$. This step profile is determined via the method in \cite{Chaplain2022eSPP}, where an effective refractive index relates the two speeds of the incoming and converted waves through
\begin{equation}
    h = \frac{2\pi m}{k_i(\tilde{n}-1)},
    \label{eq:hs}
\end{equation}
where $m$ is again the modal index of the desired flexural mode ($m =3$, here) and $\tilde{n} = c_f/c_i$ is the ratio of the wavespeeds of the converted flexural and incident longitudinal waves respectively, with $k_i$ the wavenumber of the incident mode along the pipe axis. To reduce the length of the step size the spiral is partitioned into three turns such that $h_s = h/3$. The wavespeeds of each mode are determined through the dispersion of the pipe, evaluated by spectral collocation \cite{adamou2004spectral,Chaplain2022eSPP}.

\section*{Transfer of Elastic OAM} \label{sec:transfer}

\label{sec:methods}

\begin{figure}
    \centering
    \includegraphics[width = 0.495\textwidth]{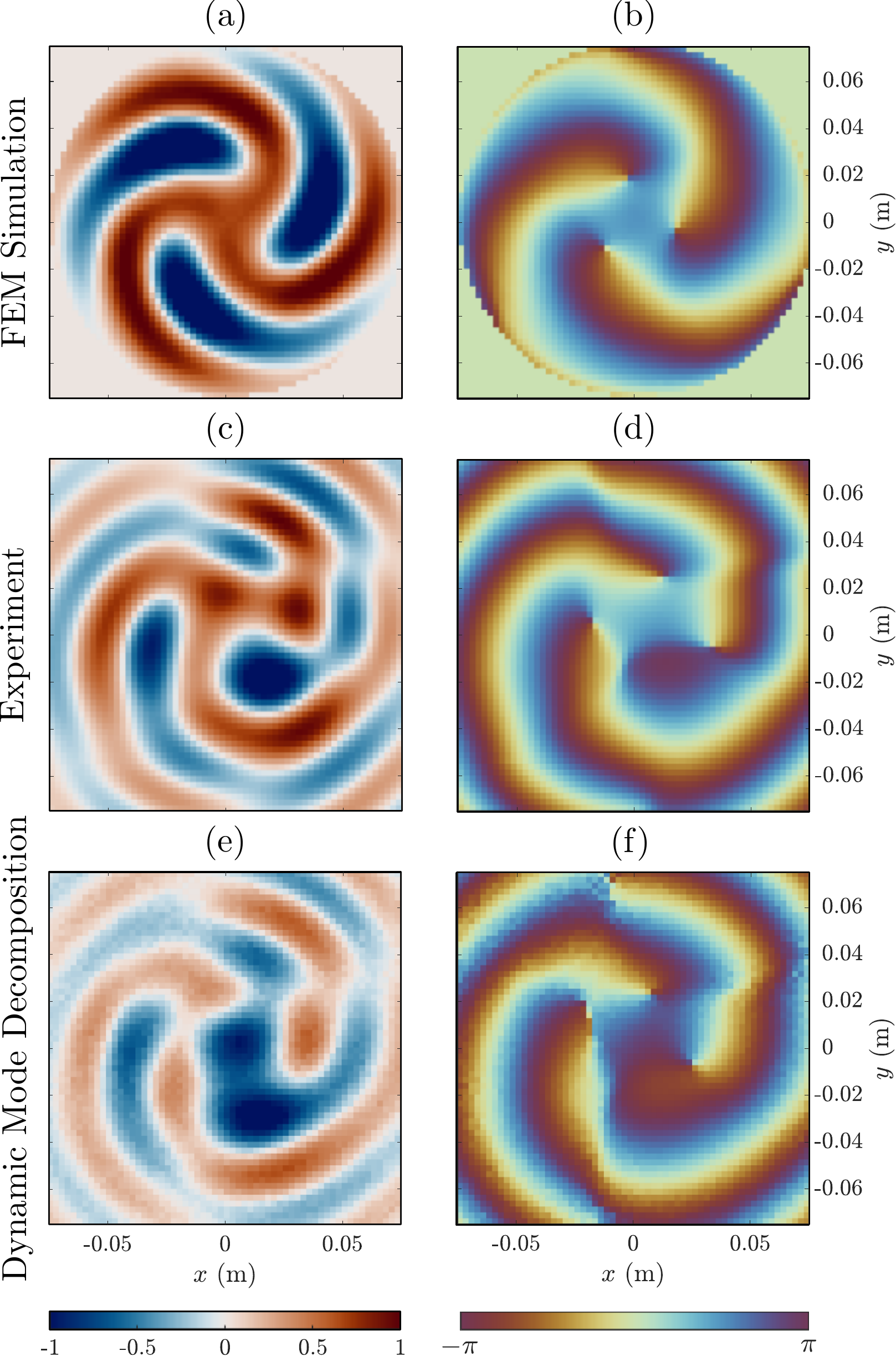}
    \caption{Results and Comparisons: Frequency domain real pressure field (left column) and phase (right column), at $58~\si{\kilo\hertz}$. The measurement plane is in the fluid at a depth $20~\si{\milli\meter}$ from the submerged end of the pipe for (a-b) FEM Simulation, (c-d) Experiment and (e-f) Dynamic Mode Decomposition (performed on experimental data set - see Appendix~\ref{sec:apB}).}
    \label{fig:Results}
\end{figure}

To experimentally confirm the transfer of elastic OAM, we consider the coupling of the compressional component of a guided flexural $F(3,2)$ mode in an elastic pipe, to the acoustic pressure field in a fluid (water) in which the pipe is partially submerged. Extensive time-gated acoustic characterisation of the fluid-field pressure distributions were made using a scanning tank facility, shown in Fig.~\ref{fig:Pipe}(b-d). We show, in Figure~\ref{fig:Results}, the first experimental observation of elastic orbital angular momentum transfer by an elastic spiral phase pipe. The time-series data obtained (see Appendix~\ref{sec:apA}) is analysed by way of the Fast Fourier Transform (FFT), giving the spatial-frequency components comprising the acoustic signal in the fluid. Figures~\ref{fig:Results}(a-b) show the real pressure field and the phase, respectively, of the FFT of FEM simulations (see Appendix~\ref{sec:apC}) at $58~\si{\kilo\hertz}$, $20~\si{\milli\meter}$ below the submerged end of the pipe. The discrepancy between the design frequency and the shown frequency arise due to imperfections in the eSPP milling procedure, as a consequence of the finite size of the drill head, as well as the effect of real damping and viscous properties of the materials at high frequencies. The result is that the spiral tips are rounded meaning the step profile is not completely accurate for the design frequency; the device is not purely monochromatic and works over a range of frequencies near the design frequency, with varying efficiency \cite{Chaplain2022eSPP}. The corresponding experimental pressure fields are shown in Figs.~\ref{fig:Results}(c-d); there is clear qualitative agreement between simulation and experiment: we observe the predicted triple-helix phase profile with three phase singularities in the form of acoustic vortices. The splitting of the central vortex into three first order charges results from the instability of higher order OAM modes \cite{basistiy1993optics}. Similar to optical beams, these modes are vulnerable to perturbation by any coherent background \cite{ricci2012instability}, that itself does not require any dislocation lines (vortices) \cite{berry2001knotted}; unconverted compressional waves form such a background resulting in the observed decay of the high-order screw dislocations on a sum of dislocations of charge one. These are clearly visible in the simulations and experimental data. Additionally in the experiment there is an amplitude modulation due to the physical eSPP only approximating the exactly circular-helicoid structure. 

\begin{figure}
    \centering
    \includegraphics[width = 0.45\textwidth]{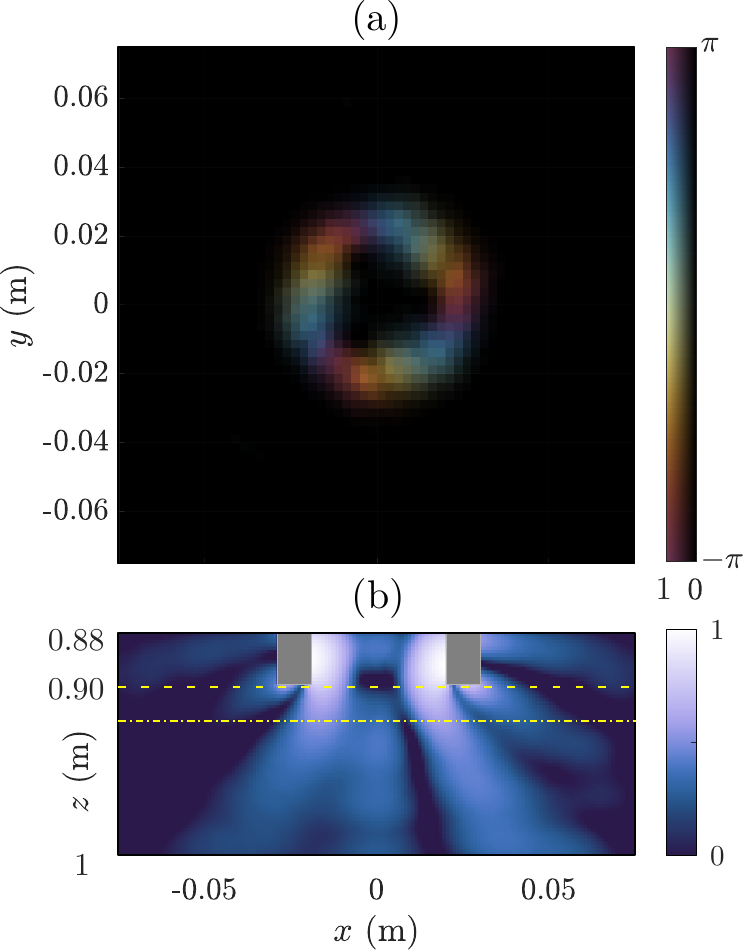}
    \caption{Effect of Measurement Depth (FEM simulation): Fourier analysed time domain simulation with (a) Intensity (hue) and phase (colour) showing typical ``doughnut'' acoustic beam indicative of OAM at a plane close to the pipe end ($1~\si{\milli\meter}$ below). (b) Normalised acoustic Intensity along $z-x$ plane ($y = 0$): data extracted along the dashed yellow line is used in (a) and the dash-dotted line in Fig.~\ref{fig:Results}(a-b) ($20~\si{\milli\meter}$ below the pipe). The acoustic vortex is obscured at the lower plane due to interaction with other modes present in the pipe, the cross-section of which is shown by the grey rectangles.}
    \label{fig:LG}
\end{figure}

Figure~\ref{fig:Pipe}(b) shows a zoom of the measurement area, with example experimental result superimposed, highlighting that due to both the parallax associated with refraction in the alignment, and the physical dimensions of the hydrophone, only an approximate depth of the relative position of the acoustic centre of the hydrophone (i.e. the plane where the pressure is accurately mapped) can be determined. The matching of the field profiles is observed at a depth of $20~\si{\milli\meter}$ below the pipe. This is, approximately, the closest possible approach of the acoustic centre of the hydrophone. The implications of this are highlighted in Fig.~\ref{fig:LG}.
For an acoustic Bessel-like beam carrying OAM, one expects a zero in acoustic intensity due to the phase singularity at the centre of the beam, where the intensity in given by
\begin{equation}
    \boldsymbol{I} = \frac{1}{4}\left(p\boldsymbol{v}^{*} + p^{*}\boldsymbol{v} \right),
\end{equation}
where, in the frequency domain, the velocity, $\boldsymbol{v}$ is related to the pressure $p$ through $\boldsymbol{v} = \frac{-1}{i\omega\rho}\nabla p$ and $*$ denotes the complex conjugate. At the observable depth this is obscured due to the background field excited from the compressional mode that is unconverted by the eSPP, since it is not perfectly efficient \cite{Chaplain2022eSPP}. As such at the measurement plane, marked by the dash-dotted line in Fig.~\ref{fig:LG}(b), there is an amplitude modulation of the Bessel-like nature resulting from modal interference. However, close to the end of the pipe, e.g. at the plane marked by the dashed line in Fig.~\ref{fig:LG}(b), the doughnut-like profile of the beam is unperturbed; this is seen in Fig.~\ref{fig:LG}(a) that shows the Fourier-analysed FEM acoustic intensity and phase as a hued colourmap. 

To confirm that the observed pressure field in the fluid is the dominant mode within the system, despite the pipe being modally rich, we perform Dynamic Mode Decomposition (DMD) on the experimental data set (see Appendix~\ref{sec:apB}). This is a technique popularised by Schmid \cite{schmid_2010} that extracts the singular values of a matrix representing the time-evolution of the complete data set, and thus determines the dominant dynamics of the system. In Figs.~\ref{fig:Results}(e-f) we show the results of the DMD on the experimental data, corroborating the assertion that the helical pressure field propagating through the fluid is dominant, as a direct result of the coupling from the compressional component of the incident $F(3,2)$ mode designed to be excited by the eSPP.

We further explain the amplitude variation of the pressure field in the fluid by considering the superposition of the OAM beam with background sources. We do so by an analogy to amplitude-modulated-discrete-phased acoustic sources that are conventionally used for exciting acoustic beam shapes that carry OAM.

\subsection*{Analogy to Discrete Phased Arrays}
\label{sec:discrete}
\begin{figure*}
    \centering
    \includegraphics[width = 0.85\textwidth]{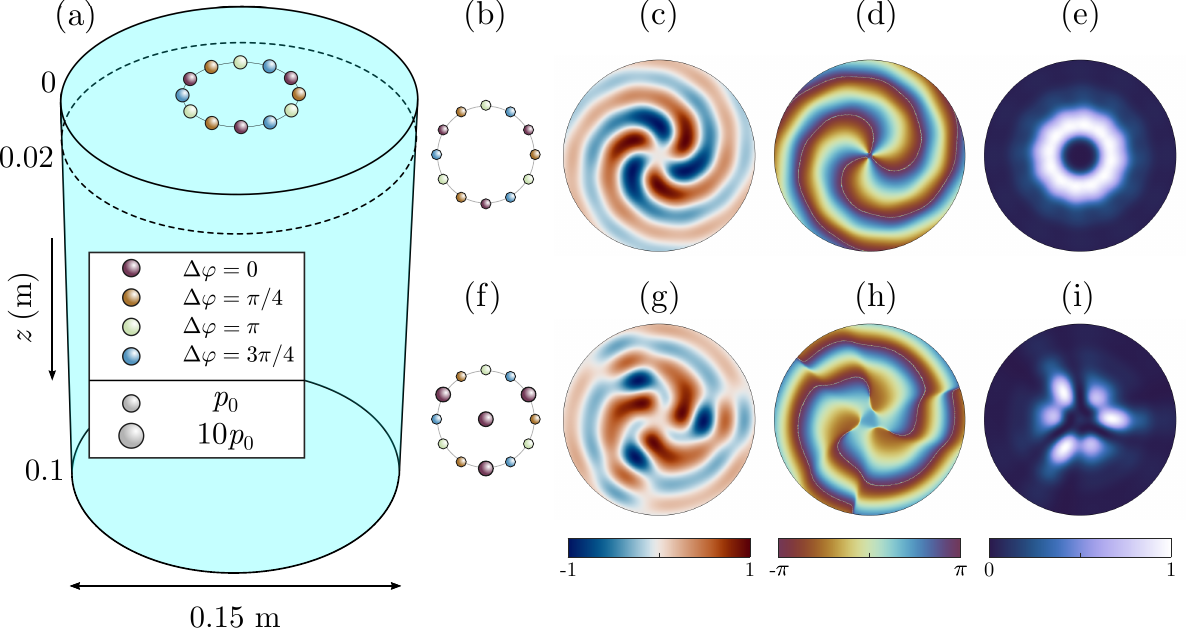}
    \caption{Analogy to Discrete Phased Acoustic Sources: (a) FEM geometry showing ring of 12 point acoustic sources (coloured spheres), with relative phase and amplitude shown by their colour and size respectively, atop cylindrical region of water with cylindrical radiation conditions on all boundaries. (b) geometry of case (i) for conventional approximation of acoustic OAM beams; real pressure, phase, and acoustic intensity are shown in (c,d,e) respectively for this case at a depth of $20~\si{\milli\meter}$ below the source position. (f) geometry of case (ii), where amplitude modulation is incorporated by an additional central source and by altering the amplitudes of three of the sources around the ring. (g,h,i) are analogous to (c,d,e) respectively for case (ii), showing the amplitude modulation of the acoustic OAM beam.}
    \label{fig:discrete}
\end{figure*}

Conventional methods for exciting acoustic beams that carry orbital angular momentum rely on discrete phased sources, such as circular arrays of loudspeakers. The interference of the monopolar-like sources then approximates a beam with a helical phase-front. Often acoustic waveguides are used to enable the beam waist to be formed a desired distance away from the sources \cite{gibson2018reversal,cromb2020amplification}. In Fig.~\ref{fig:discrete} we show the discrete phased analogy via a frequency domain FEM simulation for two cases; (i) a ring of 12 phased point acoustic sources with equal amplitude, and (ii) the same ring but with additional central source and amplitude variation. We included this model purely as a qualitative analogue to the acoustic OAM mode excited by the pipe; the vector elastic system cannot be reduced to a purely scalar (monopole) source as higher order vector components (e.g. dipole) contribute \cite{bruus2012acoustofluidics,toftul2019acoustic}. The geometry considered is such that the point sources lie on a ring of diameter $50~\si{\milli\meter}$, as if placed at the mid-point of the pipe thickness, lying atop a cylindrical volume of water $0.1~\si{meter}$ deep and $0.15~\si{meter}$ in diameter, akin to the FEM simulations of the main experiment (see see Appendix~\ref{sec:apC}). Each source is coloured to represent the relative phase shift (of $\pi/4$ radians) to the adjacent sources, and has an amplitude represented by their relative size. The array is chosen so that a topological charge of $m = 3$ is achieved. Figures~\ref{fig:discrete}(c-d) show, for case (i), the real pressure field, phase, and acoustic intensity respectively for an excitation frequency of $58~\si{\kilo\hertz}$ at a distance $20~\si{\milli\meter}$ from the source plane. As there is no other sources present this well approximates an acoustic beam carrying OAM. For case (ii), an additional source is present at the centre of the ring, and the amplitude of three sources is also modified. 

The eSPP considered throughout can be seen to act as a continuous phased acoustic source, with the phase profile determined by the circumferential order of the flexural mode, as shown in Fig.~\ref{fig:LG}(a). The analogy to the additional source in case (ii) represents the amplitude modulation in the experiment due to the background provided by the unconverted compressional wave, and other modes present in the pipe. This intuitive analogy gives qualitative agreement to the experimental fields. As such we pose that the pipe acts as an amplitude-modulated-continuous-phased source for acoustic OAM beams. 

\section*{conclusions and perspectives} \label{sec:conc}

Generating acoustic vortex beams that carry OAM has been instrumental to the development of optical and acoustic tweezers. Specifically in acoustics, the generation of these modes conventionally relies on discrete phased arrays. By considering the elastic orbital angular momentum associated with compressional motion we have demonstrated the first experimental observation of elastic orbital angular momentum transfer from guided flexural modes in a pipe to acoustic waves in a fluid, verifying the experiment proposed in \cite{Chaplain2022eOAM}, and thus providing a new avenue to generate acoustic OAM beams. 



The applications of this phenomena are therefore aligned with those of acoustic tweezers, including sensing, communication and microfluidic control. Extensions of this new methodology are anticipated to fluid-filled pipes, where there exist attractive applications in, for example, non-destructive testing in pipe-networks.

\section*{Acknowledgements}
G.J.C gratefully acknowledges financial support from the Royal Commission for the Exhibition of 1851 in the form of a Research Fellowship. J.M.D.P acknowledges the financial support from the H2020 FET-proactive project MetaVEH under grant agreement No. 952039. T.A.S gratefully acknowledges financial support from DSTL. The authors thank Prof. R.~V. Craster, Prof. A.~P. Hibbins and Dr. S.~A.~R. Horsely for useful conversations, and to G.~T. Starkey for assistance in mount construction.

\appendix


\section{Experimental Setup}
\label{sec:apA}
Measurements were performed in a water tank without wall or surface treatments, with dimensions $3.0 \times 1.8 \times 1.2$ m ($L \times W \times D$). The pipe was suspended vertically above the tank using nylon fishing line attached to a mount so that the end of the pipe was submerged approximately $20~\si{\milli\meter}$ into the fluid (Fig.~\ref{fig:Pipe}(d)); the seal on the capped end of the pipe ensures fluid is present within the pipe, up to the same depth of submersion. A piezoelectric PZT-8 disc of thickness $12~\si{\milli\meter}$ and diameter $35~\si{\milli\meter}$ was glued to the centre of the cap to provide excitation with a 5-cycle pulse centred on $60~\si{\kilo\hertz}$. The piezoelectric excites the $L(0,2)$ mode which then efficiently excites the $F(3,2)$ mode via mode conversion in the spiral region as outlined in \cite{Chaplain2022eSPP}. 

To obtain pressure field maps of sound radiated from the submerged end of the pipe, the signal at the detection hydrophone (Brüel \& Kjær 8103 hydrophone) was scanned in space using an $xyz$ scanning stage (in-house built with Aerotech controllers). The hydrophone was vertically mounted to a perforated perspex arm, to match the propagation direction of the acoustic field. The acoustic propagation was then spatially mapped in $2.5~\si{\milli\meter}$ steps across a $75 \times 75~\si{\milli\meter}^2$ area centred beneath the pipe; the voltage, $V$, from the detector was recorded as a function of time, $t$, at each position in the scan. At each spatial point, the acoustic pressure field is averaged over the detecting area of the hydrophone head and the signals were averaged in time over 20 repeat pulses to improve the signal-to-noise ratio. The detector was sampled with sample rate $f_s = 9.62~\si{\mega\hertz}$ to record the signal for $5.2~\si{\milli\second}$ at each point. The resulting usable frequency range for this source-detector response function was between $26-90~\si{\kilo\hertz}$. The resulting signals are time-windowed corresponding to the time-of-flight of the flexural wave packet so that the pressure field excited by the $F(3,2)$ pulse is isolated in the fluid. We then confirm this is a dominant mode of the system through Dynamic Mode Decomposition.

\section{Dynamic Mode Decomposition} \label{sec:apB}
Dynamic Mode Decomposition is a technique developed by Schmid \cite{schmid_2010} that enables a data set, be it numerical or experimental, to be analysed so that the dominant dynamics can be observed. This is a particularly attractive method here given the large number of modes excited within the pipe. Here we briefly outline the methodology following \cite{schmid_2010}.

DMD rests on representing an original time-series data set $D$ as a sum of $n$ mode shapes associated with the radian frequency $\omega_{n}$ such that
\begin{equation}
    D = \sum_{n}\zeta_n\exp(i\omega_n t),
\end{equation}
where $\zeta_n$ is the $n^{th}$ mode shape. The data we analyse is the temporal evolution of the acoustic pressure at a series of grid-points in space. The data is rearranged into a single matrix such that each column represents on frame of the data:
\begin{equation}
    X = \left[x_1,x_2,\dots,x_N\right],
\end{equation}
Where $X$ is the complete data set and $x_i$ is the data at times $i = 1,\dots,N$. As the governing equations for the acoustic propagation are linear, the data at each time step can be related by a matrix $A$ such that 
\begin{equation}
    x_{i+1} = Ax_{i},
\end{equation}
and thus 
\begin{equation}
    X = \left[x_1, Ax_1, \dots, A^{N-1}x_1 \right].
\end{equation}
The dynamics of the system are then governed by the eigenvalues and eigenvectors of $A$, which can be approximated by several numerical methods. Here, as in Schmid's original paper \cite{schmid_2010}, we use singular value decomposition (SVD). We shall also consider the shifted matrix
\begin{equation}
    \bar{X} = AX = \left[x_2,x_3,\dots,x_{N+1}\right]; 
    \label{eq:Xbar}
\end{equation}
for sufficiently large $N$ (i.e. a long time signal) $X$ and $\bar{X}$ will have a near identical structure. By SVD, we write
\begin{equation}
    X = USV^{\dagger},
\end{equation}
where $U$ and $V$ contain the left- and right-singular vectors respectively, with the singular values along the diagonal of $S$. If the relative size of successive singular values to the first few is small, then the size of the matrices can be reduced, with the reduced forms subsequently written as e.g. $\hat{U}$. The matrix $U$ contains the so-called principal directions, that are used to rewrite the data in a new basis and define
\begin{align}
\tilde{A} = U^{T}AU.
\end{align}
Using the reduced forms, \eqref{eq:Xbar} then becomes
\begin{align}
    \begin{split}
        \bar{X} &\approx A\left(\hat{U}\hat{S}\hat{V}^{T}\right) \\
        \implies \tilde{A} &\approx \hat{U}^T\bar{X}\hat{V}\hat{S}^{-1}.
    \end{split}
\end{align}
This approximation of $A$ then contains all the information needed to take one frame of the data to the next. The eigenvalues and eigenvectors of $\tilde{A}$ then are obtained by converting back to the original basis such that 
\begin{equation}
    \zeta_n = \hat{U}\eta_{n},
\end{equation}
where $\zeta_n$ is the $n^{th}$ mode for the $n^{th}$ eigenvector $\eta_{n}$. The results of this decomposition on the time-series data obtained in the experiments is shown in Fig.~\ref{fig:Results}(e-f), showing that this is a dominant mode shape.

\section{Finite Element Modelling} \label{sec:apC}
\begin{figure}[h!]
    \centering
    \includegraphics[width = 0.35\textwidth]{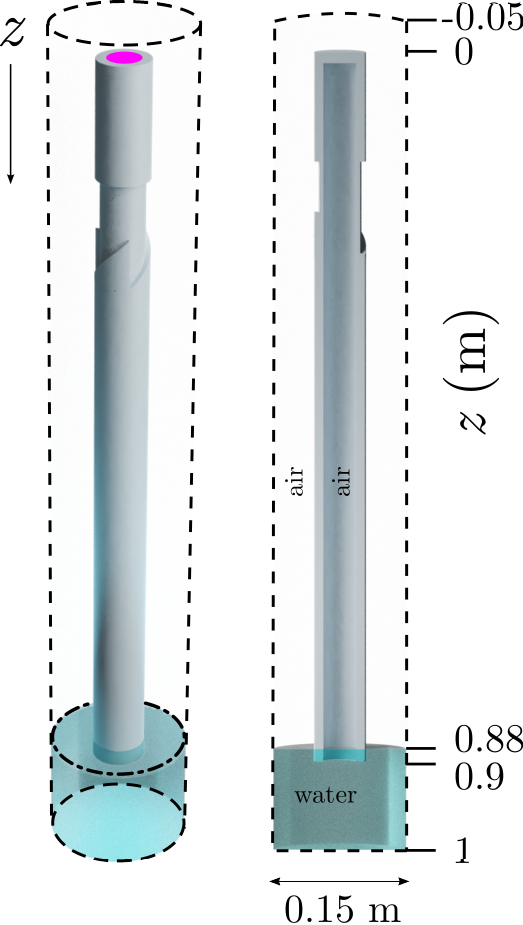}
    \caption{FEM Simulation Domain: Fully coupled acousto-elastic equations are solved for between the aluminium pipe and air, the pipe and the water, and the air and water. The boundary load applied at $z = 0$ to top surface of the pipe represents the excitation from the piezoelectric disc, shown as magenta circle of diameter $35 $ mm. Dashed lines correspond to cylindrical radiation boundaries.}
    \label{fig:fem}
\end{figure}

The commercial FEM software COMSOL Multiphysics\textsuperscript{\textregistered} was used to perform time domain simulations of the suspended pipe geometry. The acoustics and structural mechanics module were used with acoustic-solid interaction to couple the displacement field in the pipe with the acoustic pressure fields in the air and water. A schematic of the simulation domain is shown in Fig.~\ref{fig:fem}, with cylindrical wave radiation conditions on the dashed boundaries. The same 5-cycle tone burst, centred on $60~\si{\kilo\hertz}$ excitation was used and applied as a boundary load to the top cap of the pipe (area marked with magenta circle in Fig.~\ref{fig:fem}) to simulate the effect of the piezoelectic disc source (not actually modelled in the geometry).
The numerical pressure field was then extracted in the fluid, with the same spatial resolution as used in the experiment. Fourier analysis was then performed via the Fast Fourier Transform to obtain the spatial-frequency spectra, as done in the experiment. The results were analysed at several planes beneath the pipe to determine the position of the acoustic centre of the hydrophone and used to show the excitation of a LG-like acoustic beam near the submerged surface of the pipe (Fig.~\ref{fig:LG}).

For the comparison with a discrete phased acoustic array, only the acoustics module was used, with the simulation domain shown in Fig.~\ref{fig:discrete}, using monopolar-like point acoustic sources. The simulation domain here matches the region of water in the main simulation. 

\begin{thebibliography}{63}%
\makeatletter
\providecommand \@ifxundefined [1]{%
 \@ifx{#1\undefined}
}%
\providecommand \@ifnum [1]{%
 \ifnum #1\expandafter \@firstoftwo
 \else \expandafter \@secondoftwo
 \fi
}%
\providecommand \@ifx [1]{%
 \ifx #1\expandafter \@firstoftwo
 \else \expandafter \@secondoftwo
 \fi
}%
\providecommand \natexlab [1]{#1}%
\providecommand \enquote  [1]{``#1''}%
\providecommand \bibnamefont  [1]{#1}%
\providecommand \bibfnamefont [1]{#1}%
\providecommand \citenamefont [1]{#1}%
\providecommand \href@noop [0]{\@secondoftwo}%
\providecommand \href [0]{\begingroup \@sanitize@url \@href}%
\providecommand \@href[1]{\@@startlink{#1}\@@href}%
\providecommand \@@href[1]{\endgroup#1\@@endlink}%
\providecommand \@sanitize@url [0]{\catcode `\\12\catcode `\$12\catcode
  `\&12\catcode `\#12\catcode `\^12\catcode `\_12\catcode `\%12\relax}%
\providecommand \@@startlink[1]{}%
\providecommand \@@endlink[0]{}%
\providecommand \url  [0]{\begingroup\@sanitize@url \@url }%
\providecommand \@url [1]{\endgroup\@href {#1}{\urlprefix }}%
\providecommand \urlprefix  [0]{URL }%
\providecommand \Eprint [0]{\href }%
\providecommand \doibase [0]{https://doi.org/}%
\providecommand \selectlanguage [0]{\@gobble}%
\providecommand \bibinfo  [0]{\@secondoftwo}%
\providecommand \bibfield  [0]{\@secondoftwo}%
\providecommand \translation [1]{[#1]}%
\providecommand \BibitemOpen [0]{}%
\providecommand \bibitemStop [0]{}%
\providecommand \bibitemNoStop [0]{.\EOS\space}%
\providecommand \EOS [0]{\spacefactor3000\relax}%
\providecommand \BibitemShut  [1]{\csname bibitem#1\endcsname}%
\let\auto@bib@innerbib\@empty
\bibitem [{\citenamefont {Shen}\ \emph {et~al.}(2019)\citenamefont {Shen},
  \citenamefont {Wang}, \citenamefont {Xie}, \citenamefont {Min}, \citenamefont
  {Fu}, \citenamefont {Liu}, \citenamefont {Gong},\ and\ \citenamefont
  {Yuan}}]{shen2019optical}%
  \BibitemOpen
  \bibfield  {author} {\bibinfo {author} {\bibfnamefont {Y.}~\bibnamefont
  {Shen}}, \bibinfo {author} {\bibfnamefont {X.}~\bibnamefont {Wang}}, \bibinfo
  {author} {\bibfnamefont {Z.}~\bibnamefont {Xie}}, \bibinfo {author}
  {\bibfnamefont {C.}~\bibnamefont {Min}}, \bibinfo {author} {\bibfnamefont
  {X.}~\bibnamefont {Fu}}, \bibinfo {author} {\bibfnamefont {Q.}~\bibnamefont
  {Liu}}, \bibinfo {author} {\bibfnamefont {M.}~\bibnamefont {Gong}},\ and\
  \bibinfo {author} {\bibfnamefont {X.}~\bibnamefont {Yuan}},\ }\bibfield
  {title} {\bibinfo {title} {Optical vortices 30 years on: Oam manipulation
  from topological charge to multiple singularities},\ }\href@noop {}
  {\bibfield  {journal} {\bibinfo  {journal} {Light: Science \& Applications}\
  }\textbf {\bibinfo {volume} {8}},\ \bibinfo {pages} {1} (\bibinfo {year}
  {2019})}\BibitemShut {NoStop}%
\bibitem [{\citenamefont {Allen}\ \emph {et~al.}(1992)\citenamefont {Allen},
  \citenamefont {Beijersbergen}, \citenamefont {Spreeuw},\ and\ \citenamefont
  {Woerdman}}]{allen1992orbital}%
  \BibitemOpen
  \bibfield  {author} {\bibinfo {author} {\bibfnamefont {L.}~\bibnamefont
  {Allen}}, \bibinfo {author} {\bibfnamefont {M.~W.}\ \bibnamefont
  {Beijersbergen}}, \bibinfo {author} {\bibfnamefont {R.}~\bibnamefont
  {Spreeuw}},\ and\ \bibinfo {author} {\bibfnamefont {J.}~\bibnamefont
  {Woerdman}},\ }\bibfield  {title} {\bibinfo {title} {Orbital angular momentum
  of light and the transformation of {L}aguerre-{G}aussian laser modes},\
  }\href@noop {} {\bibfield  {journal} {\bibinfo  {journal} {Phys. Rev. A}\
  }\textbf {\bibinfo {volume} {45}},\ \bibinfo {pages} {8185} (\bibinfo {year}
  {1992})}\BibitemShut {NoStop}%
\bibitem [{\citenamefont {Fontaine}\ \emph {et~al.}(2019)\citenamefont
  {Fontaine}, \citenamefont {Ryf}, \citenamefont {Chen}, \citenamefont
  {Neilson}, \citenamefont {Kim},\ and\ \citenamefont
  {Carpenter}}]{fontaine2019laguerre}%
  \BibitemOpen
  \bibfield  {author} {\bibinfo {author} {\bibfnamefont {N.~K.}\ \bibnamefont
  {Fontaine}}, \bibinfo {author} {\bibfnamefont {R.}~\bibnamefont {Ryf}},
  \bibinfo {author} {\bibfnamefont {H.}~\bibnamefont {Chen}}, \bibinfo {author}
  {\bibfnamefont {D.~T.}\ \bibnamefont {Neilson}}, \bibinfo {author}
  {\bibfnamefont {K.}~\bibnamefont {Kim}},\ and\ \bibinfo {author}
  {\bibfnamefont {J.}~\bibnamefont {Carpenter}},\ }\bibfield  {title} {\bibinfo
  {title} {Laguerre-gaussian mode sorter},\ }\href@noop {} {\bibfield
  {journal} {\bibinfo  {journal} {Nat. Comms.}\ }\textbf {\bibinfo {volume}
  {10}},\ \bibinfo {pages} {1} (\bibinfo {year} {2019})}\BibitemShut {NoStop}%
\bibitem [{\citenamefont {Yao}\ and\ \citenamefont
  {Padgett}(2011)}]{yao2011orbital}%
  \BibitemOpen
  \bibfield  {author} {\bibinfo {author} {\bibfnamefont {A.~M.}\ \bibnamefont
  {Yao}}\ and\ \bibinfo {author} {\bibfnamefont {M.~J.}\ \bibnamefont
  {Padgett}},\ }\bibfield  {title} {\bibinfo {title} {Orbital angular momentum:
  origins, behavior and applications},\ }\href@noop {} {\bibfield  {journal}
  {\bibinfo  {journal} {Adv. Opt. Photonics}\ }\textbf {\bibinfo {volume}
  {3}},\ \bibinfo {pages} {161} (\bibinfo {year} {2011})}\BibitemShut {NoStop}%
\bibitem [{\citenamefont {Loke}\ \emph {et~al.}(2014)\citenamefont {Loke},
  \citenamefont {Asavei}, \citenamefont {Stilgoe}, \citenamefont {Nieminen},\
  and\ \citenamefont {Rubinsztein-Dunlop}}]{loke2014driving}%
  \BibitemOpen
  \bibfield  {author} {\bibinfo {author} {\bibfnamefont {V.~L.}\ \bibnamefont
  {Loke}}, \bibinfo {author} {\bibfnamefont {T.}~\bibnamefont {Asavei}},
  \bibinfo {author} {\bibfnamefont {A.~B.}\ \bibnamefont {Stilgoe}}, \bibinfo
  {author} {\bibfnamefont {T.~A.}\ \bibnamefont {Nieminen}},\ and\ \bibinfo
  {author} {\bibfnamefont {H.}~\bibnamefont {Rubinsztein-Dunlop}},\ }\bibfield
  {title} {\bibinfo {title} {Driving corrugated donut rotors with
  laguerre-gauss beams},\ }\href@noop {} {\bibfield  {journal} {\bibinfo
  {journal} {Opt. Express}\ }\textbf {\bibinfo {volume} {22}},\ \bibinfo
  {pages} {19692} (\bibinfo {year} {2014})}\BibitemShut {NoStop}%
\bibitem [{\citenamefont {Allen}\ \emph {et~al.}(2016)\citenamefont {Allen},
  \citenamefont {Barnett},\ and\ \citenamefont {Padgett}}]{allen2016optical}%
  \BibitemOpen
  \bibfield  {author} {\bibinfo {author} {\bibfnamefont {L.}~\bibnamefont
  {Allen}}, \bibinfo {author} {\bibfnamefont {S.~M.}\ \bibnamefont {Barnett}},\
  and\ \bibinfo {author} {\bibfnamefont {M.~J.}\ \bibnamefont {Padgett}},\
  }\href@noop {} {\emph {\bibinfo {title} {Optical angular momentum}}}\
  (\bibinfo  {publisher} {CRC press},\ \bibinfo {year} {2016})\BibitemShut
  {NoStop}%
\bibitem [{\citenamefont {Padgett}(2017)}]{padgett2017orbital}%
  \BibitemOpen
  \bibfield  {author} {\bibinfo {author} {\bibfnamefont {M.~J.}\ \bibnamefont
  {Padgett}},\ }\bibfield  {title} {\bibinfo {title} {Orbital angular momentum
  25 years on},\ }\href@noop {} {\bibfield  {journal} {\bibinfo  {journal}
  {Opt. Express}\ }\textbf {\bibinfo {volume} {25}},\ \bibinfo {pages} {11265}
  (\bibinfo {year} {2017})}\BibitemShut {NoStop}%
\bibitem [{\citenamefont {Barnett}\ \emph {et~al.}(2017)\citenamefont
  {Barnett}, \citenamefont {Babiker},\ and\ \citenamefont
  {Padgett}}]{barnett2017optical}%
  \BibitemOpen
  \bibfield  {author} {\bibinfo {author} {\bibfnamefont {S.~M.}\ \bibnamefont
  {Barnett}}, \bibinfo {author} {\bibfnamefont {M.}~\bibnamefont {Babiker}},\
  and\ \bibinfo {author} {\bibfnamefont {M.~J.}\ \bibnamefont {Padgett}},\
  }\bibfield  {title} {\bibinfo {title} {Optical orbital angular momentum},\
  }\href {https://doi.org/10.1098/rsta.2015.0444} {\bibfield  {journal}
  {\bibinfo  {journal} {Philos. Trans. R. Soc. A}\ }\textbf {\bibinfo {volume}
  {375}},\ \bibinfo {pages} {20150444} (\bibinfo {year} {2017})}\BibitemShut
  {NoStop}%
\bibitem [{\citenamefont {Higgins}(1992)}]{LaserFocusWorld}%
  \BibitemOpen
  \bibfield  {author} {\bibinfo {author} {\bibfnamefont {T.~V.}\ \bibnamefont
  {Higgins}},\ }\bibfield  {title} {\bibinfo {title} {Spiral waveplate design
  produces radially polarized laser light},\ }\href@noop {} {\bibfield
  {journal} {\bibinfo  {journal} {Laser Focus World}\ }\textbf {\bibinfo
  {volume} {28}},\ \bibinfo {pages} {18} (\bibinfo {year} {1992})}\BibitemShut
  {NoStop}%
\bibitem [{\citenamefont {Oemrawsingh}\ \emph {et~al.}(2004)\citenamefont
  {Oemrawsingh}, \citenamefont {Van~Houwelingen}, \citenamefont {Eliel},
  \citenamefont {Woerdman}, \citenamefont {Verstegen}, \citenamefont
  {Kloosterboer} \emph {et~al.}}]{oemrawsingh2004production}%
  \BibitemOpen
  \bibfield  {author} {\bibinfo {author} {\bibfnamefont {S.}~\bibnamefont
  {Oemrawsingh}}, \bibinfo {author} {\bibfnamefont {J.}~\bibnamefont
  {Van~Houwelingen}}, \bibinfo {author} {\bibfnamefont {E.}~\bibnamefont
  {Eliel}}, \bibinfo {author} {\bibfnamefont {J.}~\bibnamefont {Woerdman}},
  \bibinfo {author} {\bibfnamefont {E.}~\bibnamefont {Verstegen}}, \bibinfo
  {author} {\bibfnamefont {J.}~\bibnamefont {Kloosterboer}}, \emph {et~al.},\
  }\bibfield  {title} {\bibinfo {title} {Production and characterization of
  spiral phase plates for optical wavelengths},\ }\href@noop {} {\bibfield
  {journal} {\bibinfo  {journal} {Appl. Opt.}\ }\textbf {\bibinfo {volume}
  {43}},\ \bibinfo {pages} {688} (\bibinfo {year} {2004})}\BibitemShut
  {NoStop}%
\bibitem [{\citenamefont {O'Neil}\ \emph {et~al.}(2002)\citenamefont {O'Neil},
  \citenamefont {MacVicar}, \citenamefont {Allen},\ and\ \citenamefont
  {Padgett}}]{o2002intrinsic}%
  \BibitemOpen
  \bibfield  {author} {\bibinfo {author} {\bibfnamefont {A.}~\bibnamefont
  {O'Neil}}, \bibinfo {author} {\bibfnamefont {I.}~\bibnamefont {MacVicar}},
  \bibinfo {author} {\bibfnamefont {L.}~\bibnamefont {Allen}},\ and\ \bibinfo
  {author} {\bibfnamefont {M.}~\bibnamefont {Padgett}},\ }\bibfield  {title}
  {\bibinfo {title} {Intrinsic and extrinsic nature of the orbital angular
  momentum of a light beam},\ }\href@noop {} {\bibfield  {journal} {\bibinfo
  {journal} {Phys. Rev. Lett.}\ }\textbf {\bibinfo {volume} {88}},\ \bibinfo
  {pages} {053601} (\bibinfo {year} {2002})}\BibitemShut {NoStop}%
\bibitem [{\citenamefont {McGloin}\ \emph {et~al.}(2003)\citenamefont
  {McGloin}, \citenamefont {Garc{\'e}s-Ch{\'a}vez},\ and\ \citenamefont
  {Dholakia}}]{mcgloin2003interfering}%
  \BibitemOpen
  \bibfield  {author} {\bibinfo {author} {\bibfnamefont {D.}~\bibnamefont
  {McGloin}}, \bibinfo {author} {\bibfnamefont {V.}~\bibnamefont
  {Garc{\'e}s-Ch{\'a}vez}},\ and\ \bibinfo {author} {\bibfnamefont
  {K.}~\bibnamefont {Dholakia}},\ }\bibfield  {title} {\bibinfo {title}
  {Interfering bessel beams for optical micromanipulation},\ }\href@noop {}
  {\bibfield  {journal} {\bibinfo  {journal} {Opt. Lett.}\ }\textbf {\bibinfo
  {volume} {28}},\ \bibinfo {pages} {657} (\bibinfo {year} {2003})}\BibitemShut
  {NoStop}%
\bibitem [{\citenamefont {Vasilyeu}\ \emph {et~al.}(2009)\citenamefont
  {Vasilyeu}, \citenamefont {Dudley}, \citenamefont {Khilo},\ and\
  \citenamefont {Forbes}}]{vasilyeu2009generating}%
  \BibitemOpen
  \bibfield  {author} {\bibinfo {author} {\bibfnamefont {R.}~\bibnamefont
  {Vasilyeu}}, \bibinfo {author} {\bibfnamefont {A.}~\bibnamefont {Dudley}},
  \bibinfo {author} {\bibfnamefont {N.}~\bibnamefont {Khilo}},\ and\ \bibinfo
  {author} {\bibfnamefont {A.}~\bibnamefont {Forbes}},\ }\bibfield  {title}
  {\bibinfo {title} {Generating superpositions of higher--order bessel beams},\
  }\href@noop {} {\bibfield  {journal} {\bibinfo  {journal} {Opt. Express}\
  }\textbf {\bibinfo {volume} {17}},\ \bibinfo {pages} {23389} (\bibinfo {year}
  {2009})}\BibitemShut {NoStop}%
\bibitem [{\citenamefont {Leach}\ \emph {et~al.}(2004)\citenamefont {Leach},
  \citenamefont {Dennis}, \citenamefont {Courtial},\ and\ \citenamefont
  {Padgett}}]{leach2004knotted}%
  \BibitemOpen
  \bibfield  {author} {\bibinfo {author} {\bibfnamefont {J.}~\bibnamefont
  {Leach}}, \bibinfo {author} {\bibfnamefont {M.~R.}\ \bibnamefont {Dennis}},
  \bibinfo {author} {\bibfnamefont {J.}~\bibnamefont {Courtial}},\ and\
  \bibinfo {author} {\bibfnamefont {M.~J.}\ \bibnamefont {Padgett}},\
  }\bibfield  {title} {\bibinfo {title} {Knotted threads of darkness},\
  }\href@noop {} {\bibfield  {journal} {\bibinfo  {journal} {Nature}\ }\textbf
  {\bibinfo {volume} {432}},\ \bibinfo {pages} {165} (\bibinfo {year}
  {2004})}\BibitemShut {NoStop}%
\bibitem [{\citenamefont {Leach}\ \emph {et~al.}(2005)\citenamefont {Leach},
  \citenamefont {Dennis}, \citenamefont {Courtial},\ and\ \citenamefont
  {Padgett}}]{leach2005vortex}%
  \BibitemOpen
  \bibfield  {author} {\bibinfo {author} {\bibfnamefont {J.}~\bibnamefont
  {Leach}}, \bibinfo {author} {\bibfnamefont {M.~R.}\ \bibnamefont {Dennis}},
  \bibinfo {author} {\bibfnamefont {J.}~\bibnamefont {Courtial}},\ and\
  \bibinfo {author} {\bibfnamefont {M.~J.}\ \bibnamefont {Padgett}},\
  }\bibfield  {title} {\bibinfo {title} {Vortex knots in light},\ }\href@noop
  {} {\bibfield  {journal} {\bibinfo  {journal} {New J. Phys.}\ }\textbf
  {\bibinfo {volume} {7}},\ \bibinfo {pages} {55} (\bibinfo {year}
  {2005})}\BibitemShut {NoStop}%
\bibitem [{\citenamefont {G{\"o}tte}\ \emph {et~al.}(2008)\citenamefont
  {G{\"o}tte}, \citenamefont {O’Holleran}, \citenamefont {Preece},
  \citenamefont {Flossmann}, \citenamefont {Franke-Arnold}, \citenamefont
  {Barnett},\ and\ \citenamefont {Padgett}}]{gotte2008light}%
  \BibitemOpen
  \bibfield  {author} {\bibinfo {author} {\bibfnamefont {J.~B.}\ \bibnamefont
  {G{\"o}tte}}, \bibinfo {author} {\bibfnamefont {K.}~\bibnamefont
  {O’Holleran}}, \bibinfo {author} {\bibfnamefont {D.}~\bibnamefont
  {Preece}}, \bibinfo {author} {\bibfnamefont {F.}~\bibnamefont {Flossmann}},
  \bibinfo {author} {\bibfnamefont {S.}~\bibnamefont {Franke-Arnold}}, \bibinfo
  {author} {\bibfnamefont {S.~M.}\ \bibnamefont {Barnett}},\ and\ \bibinfo
  {author} {\bibfnamefont {M.~J.}\ \bibnamefont {Padgett}},\ }\bibfield
  {title} {\bibinfo {title} {Light beams with fractional orbital angular
  momentum and their vortex structure},\ }\href@noop {} {\bibfield  {journal}
  {\bibinfo  {journal} {Opt. Express}\ }\textbf {\bibinfo {volume} {16}},\
  \bibinfo {pages} {993} (\bibinfo {year} {2008})}\BibitemShut {NoStop}%
\bibitem [{\citenamefont {Morgan}\ \emph {et~al.}(2017)\citenamefont {Morgan},
  \citenamefont {Miller}, \citenamefont {Li}, \citenamefont {Li},\ and\
  \citenamefont {Johnson}}]{morgan2017higher}%
  \BibitemOpen
  \bibfield  {author} {\bibinfo {author} {\bibfnamefont {K.}~\bibnamefont
  {Morgan}}, \bibinfo {author} {\bibfnamefont {J.}~\bibnamefont {Miller}},
  \bibinfo {author} {\bibfnamefont {W.}~\bibnamefont {Li}}, \bibinfo {author}
  {\bibfnamefont {Y.}~\bibnamefont {Li}},\ and\ \bibinfo {author}
  {\bibfnamefont {E.}~\bibnamefont {Johnson}},\ }\bibfield  {title} {\bibinfo
  {title} {Higher order bessel beams integrated in time (hobbit) for free space
  underwater sensing and communication},\ }in\ \href@noop {} {\emph {\bibinfo
  {booktitle} {OCEANS 2017-Anchorage}}}\ (\bibinfo {organization} {IEEE},\
  \bibinfo {year} {2017})\ pp.\ \bibinfo {pages} {1--4}\BibitemShut {NoStop}%
\bibitem [{\citenamefont {Bliokh}\ and\ \citenamefont
  {Nori}(2019)}]{bliokh2019spin}%
  \BibitemOpen
  \bibfield  {author} {\bibinfo {author} {\bibfnamefont {K.~Y.}\ \bibnamefont
  {Bliokh}}\ and\ \bibinfo {author} {\bibfnamefont {F.}~\bibnamefont {Nori}},\
  }\bibfield  {title} {\bibinfo {title} {Spin and orbital angular momenta of
  acoustic beams},\ }\href@noop {} {\bibfield  {journal} {\bibinfo  {journal}
  {Phys. Rev. B}\ }\textbf {\bibinfo {volume} {99}},\ \bibinfo {pages} {174310}
  (\bibinfo {year} {2019})}\BibitemShut {NoStop}%
\bibitem [{\citenamefont {Zhang}\ and\ \citenamefont
  {Marston}(2011)}]{zhang2011angular}%
  \BibitemOpen
  \bibfield  {author} {\bibinfo {author} {\bibfnamefont {L.}~\bibnamefont
  {Zhang}}\ and\ \bibinfo {author} {\bibfnamefont {P.~L.}\ \bibnamefont
  {Marston}},\ }\bibfield  {title} {\bibinfo {title} {Angular momentum flux of
  nonparaxial acoustic vortex beams and torques on axisymmetric objects},\
  }\href@noop {} {\bibfield  {journal} {\bibinfo  {journal} {Phys. Rev. E}\
  }\textbf {\bibinfo {volume} {84}},\ \bibinfo {pages} {065601} (\bibinfo
  {year} {2011})}\BibitemShut {NoStop}%
\bibitem [{\citenamefont {Zhang}(2018)}]{zhang2018reversals}%
  \BibitemOpen
  \bibfield  {author} {\bibinfo {author} {\bibfnamefont {L.}~\bibnamefont
  {Zhang}},\ }\bibfield  {title} {\bibinfo {title} {Reversals of orbital
  angular momentum transfer and radiation torque},\ }\href@noop {} {\bibfield
  {journal} {\bibinfo  {journal} {Phys. Rev. Appl.}\ }\textbf {\bibinfo
  {volume} {10}},\ \bibinfo {pages} {034039} (\bibinfo {year}
  {2018})}\BibitemShut {NoStop}%
\bibitem [{\citenamefont {Hong}\ \emph {et~al.}(2015)\citenamefont {Hong},
  \citenamefont {Zhang},\ and\ \citenamefont
  {Drinkwater}}]{hong2015observation}%
  \BibitemOpen
  \bibfield  {author} {\bibinfo {author} {\bibfnamefont {Z.}~\bibnamefont
  {Hong}}, \bibinfo {author} {\bibfnamefont {J.}~\bibnamefont {Zhang}},\ and\
  \bibinfo {author} {\bibfnamefont {B.~W.}\ \bibnamefont {Drinkwater}},\
  }\bibfield  {title} {\bibinfo {title} {Observation of orbital angular
  momentum transfer from bessel-shaped acoustic vortices to diphasic
  liquid-microparticle mixtures},\ }\href@noop {} {\bibfield  {journal}
  {\bibinfo  {journal} {Physical review letters}\ }\textbf {\bibinfo {volume}
  {114}},\ \bibinfo {pages} {214301} (\bibinfo {year} {2015})}\BibitemShut
  {NoStop}%
\bibitem [{\citenamefont {Baresch}\ \emph {et~al.}(2016)\citenamefont
  {Baresch}, \citenamefont {Thomas},\ and\ \citenamefont
  {Marchiano}}]{baresch2016observation}%
  \BibitemOpen
  \bibfield  {author} {\bibinfo {author} {\bibfnamefont {D.}~\bibnamefont
  {Baresch}}, \bibinfo {author} {\bibfnamefont {J.-L.}\ \bibnamefont
  {Thomas}},\ and\ \bibinfo {author} {\bibfnamefont {R.}~\bibnamefont
  {Marchiano}},\ }\bibfield  {title} {\bibinfo {title} {Observation of a
  single-beam gradient force acoustical trap for elastic particles: acoustical
  tweezers},\ }\href@noop {} {\bibfield  {journal} {\bibinfo  {journal}
  {Physical review letters}\ }\textbf {\bibinfo {volume} {116}},\ \bibinfo
  {pages} {024301} (\bibinfo {year} {2016})}\BibitemShut {NoStop}%
\bibitem [{\citenamefont {Shi}\ \emph {et~al.}(2017)\citenamefont {Shi},
  \citenamefont {Dubois}, \citenamefont {Wang},\ and\ \citenamefont
  {Zhang}}]{shi2017high}%
  \BibitemOpen
  \bibfield  {author} {\bibinfo {author} {\bibfnamefont {C.}~\bibnamefont
  {Shi}}, \bibinfo {author} {\bibfnamefont {M.}~\bibnamefont {Dubois}},
  \bibinfo {author} {\bibfnamefont {Y.}~\bibnamefont {Wang}},\ and\ \bibinfo
  {author} {\bibfnamefont {X.}~\bibnamefont {Zhang}},\ }\bibfield  {title}
  {\bibinfo {title} {High-speed acoustic communication by multiplexing orbital
  angular momentum},\ }\href@noop {} {\bibfield  {journal} {\bibinfo  {journal}
  {Proc. Natl. Acad. Sci.}\ }\textbf {\bibinfo {volume} {114}},\ \bibinfo
  {pages} {7250} (\bibinfo {year} {2017})}\BibitemShut {NoStop}%
\bibitem [{\citenamefont {Marzo}\ \emph {et~al.}(2018)\citenamefont {Marzo},
  \citenamefont {Caleap},\ and\ \citenamefont
  {Drinkwater}}]{marzo2018acoustic}%
  \BibitemOpen
  \bibfield  {author} {\bibinfo {author} {\bibfnamefont {A.}~\bibnamefont
  {Marzo}}, \bibinfo {author} {\bibfnamefont {M.}~\bibnamefont {Caleap}},\ and\
  \bibinfo {author} {\bibfnamefont {B.~W.}\ \bibnamefont {Drinkwater}},\
  }\bibfield  {title} {\bibinfo {title} {Acoustic virtual vortices with tunable
  orbital angular momentum for trapping of mie particles},\ }\href@noop {}
  {\bibfield  {journal} {\bibinfo  {journal} {Physical review letters}\
  }\textbf {\bibinfo {volume} {120}},\ \bibinfo {pages} {044301} (\bibinfo
  {year} {2018})}\BibitemShut {NoStop}%
\bibitem [{\citenamefont {Hefner}\ and\ \citenamefont
  {Marston}(1998)}]{hefner1998acoustical}%
  \BibitemOpen
  \bibfield  {author} {\bibinfo {author} {\bibfnamefont {B.~T.}\ \bibnamefont
  {Hefner}}\ and\ \bibinfo {author} {\bibfnamefont {P.~L.}\ \bibnamefont
  {Marston}},\ }\bibfield  {title} {\bibinfo {title} {Acoustical helicoidal
  waves and laguerre-gaussian beams: Applications to scattering and to angular
  momentum transport},\ }\href@noop {} {\bibfield  {journal} {\bibinfo
  {journal} {J. Acoust. Soc. Am.}\ }\textbf {\bibinfo {volume} {103}},\
  \bibinfo {pages} {2971} (\bibinfo {year} {1998})}\BibitemShut {NoStop}%
\bibitem [{\citenamefont {Hefner}\ and\ \citenamefont
  {Marston}(1999)}]{hefner1999acoustical}%
  \BibitemOpen
  \bibfield  {author} {\bibinfo {author} {\bibfnamefont {B.~T.}\ \bibnamefont
  {Hefner}}\ and\ \bibinfo {author} {\bibfnamefont {P.~L.}\ \bibnamefont
  {Marston}},\ }\bibfield  {title} {\bibinfo {title} {An acoustical helicoidal
  wave transducer with applications for the alignment of ultrasonic and
  underwater systems},\ }\href@noop {} {\bibfield  {journal} {\bibinfo
  {journal} {J. Acoust. Soc. Am.}\ }\textbf {\bibinfo {volume} {106}},\
  \bibinfo {pages} {3313} (\bibinfo {year} {1999})}\BibitemShut {NoStop}%
\bibitem [{\citenamefont {Jiang}\ \emph {et~al.}(2016)\citenamefont {Jiang},
  \citenamefont {Li}, \citenamefont {Liang}, \citenamefont {Cheng},\ and\
  \citenamefont {Zhang}}]{jiang2016convert}%
  \BibitemOpen
  \bibfield  {author} {\bibinfo {author} {\bibfnamefont {X.}~\bibnamefont
  {Jiang}}, \bibinfo {author} {\bibfnamefont {Y.}~\bibnamefont {Li}}, \bibinfo
  {author} {\bibfnamefont {B.}~\bibnamefont {Liang}}, \bibinfo {author}
  {\bibfnamefont {J.-c.}\ \bibnamefont {Cheng}},\ and\ \bibinfo {author}
  {\bibfnamefont {L.}~\bibnamefont {Zhang}},\ }\bibfield  {title} {\bibinfo
  {title} {Convert acoustic resonances to orbital angular momentum},\
  }\href@noop {} {\bibfield  {journal} {\bibinfo  {journal} {Phys. Rev. Lett.}\
  }\textbf {\bibinfo {volume} {117}},\ \bibinfo {pages} {034301} (\bibinfo
  {year} {2016})}\BibitemShut {NoStop}%
\bibitem [{\citenamefont {Naify}\ \emph {et~al.}(2016)\citenamefont {Naify},
  \citenamefont {Rohde}, \citenamefont {Martin}, \citenamefont {Nicholas},
  \citenamefont {Guild},\ and\ \citenamefont {Orris}}]{naify2016generation}%
  \BibitemOpen
  \bibfield  {author} {\bibinfo {author} {\bibfnamefont {C.~J.}\ \bibnamefont
  {Naify}}, \bibinfo {author} {\bibfnamefont {C.~A.}\ \bibnamefont {Rohde}},
  \bibinfo {author} {\bibfnamefont {T.~P.}\ \bibnamefont {Martin}}, \bibinfo
  {author} {\bibfnamefont {M.}~\bibnamefont {Nicholas}}, \bibinfo {author}
  {\bibfnamefont {M.~D.}\ \bibnamefont {Guild}},\ and\ \bibinfo {author}
  {\bibfnamefont {G.~J.}\ \bibnamefont {Orris}},\ }\bibfield  {title} {\bibinfo
  {title} {Generation of topologically diverse acoustic vortex beams using a
  compact metamaterial aperture},\ }\href@noop {} {\bibfield  {journal}
  {\bibinfo  {journal} {Appl. Phys. Lett.}\ }\textbf {\bibinfo {volume}
  {108}},\ \bibinfo {pages} {223503} (\bibinfo {year} {2016})}\BibitemShut
  {NoStop}%
\bibitem [{\citenamefont {Guo}\ \emph {et~al.}(2019)\citenamefont {Guo},
  \citenamefont {Liu}, \citenamefont {Zhou}, \citenamefont {Zhou},
  \citenamefont {Wang}, \citenamefont {Shen}, \citenamefont {Gong},
  \citenamefont {Gao}, \citenamefont {Liu},\ and\ \citenamefont
  {Guo}}]{guo2019high}%
  \BibitemOpen
  \bibfield  {author} {\bibinfo {author} {\bibfnamefont {Z.}~\bibnamefont
  {Guo}}, \bibinfo {author} {\bibfnamefont {H.}~\bibnamefont {Liu}}, \bibinfo
  {author} {\bibfnamefont {H.}~\bibnamefont {Zhou}}, \bibinfo {author}
  {\bibfnamefont {K.}~\bibnamefont {Zhou}}, \bibinfo {author} {\bibfnamefont
  {S.}~\bibnamefont {Wang}}, \bibinfo {author} {\bibfnamefont {F.}~\bibnamefont
  {Shen}}, \bibinfo {author} {\bibfnamefont {Y.}~\bibnamefont {Gong}}, \bibinfo
  {author} {\bibfnamefont {J.}~\bibnamefont {Gao}}, \bibinfo {author}
  {\bibfnamefont {S.}~\bibnamefont {Liu}},\ and\ \bibinfo {author}
  {\bibfnamefont {K.}~\bibnamefont {Guo}},\ }\bibfield  {title} {\bibinfo
  {title} {High-order acoustic vortex field generation based on a
  metasurface},\ }\href@noop {} {\bibfield  {journal} {\bibinfo  {journal}
  {Phys. Rev. E}\ }\textbf {\bibinfo {volume} {100}},\ \bibinfo {pages}
  {053315} (\bibinfo {year} {2019})}\BibitemShut {NoStop}%
\bibitem [{\citenamefont {Gibson}\ \emph {et~al.}(2018)\citenamefont {Gibson},
  \citenamefont {Toninelli}, \citenamefont {Horsley}, \citenamefont {Spalding},
  \citenamefont {Hendry}, \citenamefont {Phillips},\ and\ \citenamefont
  {Padgett}}]{gibson2018reversal}%
  \BibitemOpen
  \bibfield  {author} {\bibinfo {author} {\bibfnamefont {G.~M.}\ \bibnamefont
  {Gibson}}, \bibinfo {author} {\bibfnamefont {E.}~\bibnamefont {Toninelli}},
  \bibinfo {author} {\bibfnamefont {S.~A.}\ \bibnamefont {Horsley}}, \bibinfo
  {author} {\bibfnamefont {G.~C.}\ \bibnamefont {Spalding}}, \bibinfo {author}
  {\bibfnamefont {E.}~\bibnamefont {Hendry}}, \bibinfo {author} {\bibfnamefont
  {D.~B.}\ \bibnamefont {Phillips}},\ and\ \bibinfo {author} {\bibfnamefont
  {M.~J.}\ \bibnamefont {Padgett}},\ }\bibfield  {title} {\bibinfo {title}
  {Reversal of orbital angular momentum arising from an extreme doppler
  shift},\ }\href@noop {} {\bibfield  {journal} {\bibinfo  {journal} {Proc.
  Natl. Acad. Sci.}\ }\textbf {\bibinfo {volume} {115}},\ \bibinfo {pages}
  {3800} (\bibinfo {year} {2018})}\BibitemShut {NoStop}%
\bibitem [{\citenamefont {Antonacci}\ \emph {et~al.}(2019)\citenamefont
  {Antonacci}, \citenamefont {Caprini},\ and\ \citenamefont
  {Ruocco}}]{antonacci2019demonstration}%
  \BibitemOpen
  \bibfield  {author} {\bibinfo {author} {\bibfnamefont {G.}~\bibnamefont
  {Antonacci}}, \bibinfo {author} {\bibfnamefont {D.}~\bibnamefont {Caprini}},\
  and\ \bibinfo {author} {\bibfnamefont {G.}~\bibnamefont {Ruocco}},\
  }\bibfield  {title} {\bibinfo {title} {Demonstration of self-healing and
  scattering resilience of acoustic bessel beams},\ }\href@noop {} {\bibfield
  {journal} {\bibinfo  {journal} {Appl. Phys. Lett.}\ }\textbf {\bibinfo
  {volume} {114}},\ \bibinfo {pages} {013502} (\bibinfo {year}
  {2019})}\BibitemShut {NoStop}%
\bibitem [{\citenamefont {Cromb}\ \emph {et~al.}(2020)\citenamefont {Cromb},
  \citenamefont {Gibson}, \citenamefont {Toninelli}, \citenamefont {Padgett},
  \citenamefont {Wright},\ and\ \citenamefont
  {Faccio}}]{cromb2020amplification}%
  \BibitemOpen
  \bibfield  {author} {\bibinfo {author} {\bibfnamefont {M.}~\bibnamefont
  {Cromb}}, \bibinfo {author} {\bibfnamefont {G.~M.}\ \bibnamefont {Gibson}},
  \bibinfo {author} {\bibfnamefont {E.}~\bibnamefont {Toninelli}}, \bibinfo
  {author} {\bibfnamefont {M.~J.}\ \bibnamefont {Padgett}}, \bibinfo {author}
  {\bibfnamefont {E.~M.}\ \bibnamefont {Wright}},\ and\ \bibinfo {author}
  {\bibfnamefont {D.}~\bibnamefont {Faccio}},\ }\bibfield  {title} {\bibinfo
  {title} {Amplification of waves from a rotating body},\ }\href@noop {}
  {\bibfield  {journal} {\bibinfo  {journal} {Nat. Phys.}\ }\textbf {\bibinfo
  {volume} {16}},\ \bibinfo {pages} {1069} (\bibinfo {year}
  {2020})}\BibitemShut {NoStop}%
\bibitem [{\citenamefont {Marchiano}\ and\ \citenamefont
  {Thomas}(2005)}]{marchiano2005synthesis}%
  \BibitemOpen
  \bibfield  {author} {\bibinfo {author} {\bibfnamefont {R.}~\bibnamefont
  {Marchiano}}\ and\ \bibinfo {author} {\bibfnamefont {J.-L.}\ \bibnamefont
  {Thomas}},\ }\bibfield  {title} {\bibinfo {title} {Synthesis and analysis of
  linear and nonlinear acoustical vortices},\ }\href@noop {} {\bibfield
  {journal} {\bibinfo  {journal} {Phys. Rev. E}\ }\textbf {\bibinfo {volume}
  {71}},\ \bibinfo {pages} {066616} (\bibinfo {year} {2005})}\BibitemShut
  {NoStop}%
\bibitem [{\citenamefont {Basistiy}\ \emph {et~al.}(1993)\citenamefont
  {Basistiy}, \citenamefont {Bazhenov}, \citenamefont {Soskin},\ and\
  \citenamefont {Vasnetsov}}]{basistiy1993optics}%
  \BibitemOpen
  \bibfield  {author} {\bibinfo {author} {\bibfnamefont {I.}~\bibnamefont
  {Basistiy}}, \bibinfo {author} {\bibfnamefont {V.~Y.}\ \bibnamefont
  {Bazhenov}}, \bibinfo {author} {\bibfnamefont {M.}~\bibnamefont {Soskin}},\
  and\ \bibinfo {author} {\bibfnamefont {M.~V.}\ \bibnamefont {Vasnetsov}},\
  }\bibfield  {title} {\bibinfo {title} {Optics of light beams with screw
  dislocations},\ }\href@noop {} {\bibfield  {journal} {\bibinfo  {journal}
  {Opt. Commum.}\ }\textbf {\bibinfo {volume} {103}},\ \bibinfo {pages} {422}
  (\bibinfo {year} {1993})}\BibitemShut {NoStop}%
\bibitem [{\citenamefont {Zambrini}\ and\ \citenamefont
  {Barnett}(2006)}]{zambrini2006quasi}%
  \BibitemOpen
  \bibfield  {author} {\bibinfo {author} {\bibfnamefont {R.}~\bibnamefont
  {Zambrini}}\ and\ \bibinfo {author} {\bibfnamefont {S.~M.}\ \bibnamefont
  {Barnett}},\ }\bibfield  {title} {\bibinfo {title} {Quasi-intrinsic angular
  momentum and the measurement of its spectrum},\ }\href@noop {} {\bibfield
  {journal} {\bibinfo  {journal} {Phys. Rev. Lett.}\ }\textbf {\bibinfo
  {volume} {96}},\ \bibinfo {pages} {113901} (\bibinfo {year}
  {2006})}\BibitemShut {NoStop}%
\bibitem [{\citenamefont {Nakane}\ and\ \citenamefont
  {Kohno}(2018)}]{nakane2018angular}%
  \BibitemOpen
  \bibfield  {author} {\bibinfo {author} {\bibfnamefont {J.~J.}\ \bibnamefont
  {Nakane}}\ and\ \bibinfo {author} {\bibfnamefont {H.}~\bibnamefont {Kohno}},\
  }\bibfield  {title} {\bibinfo {title} {Angular momentum of phonons and its
  application to single-spin relaxation},\ }\href@noop {} {\bibfield  {journal}
  {\bibinfo  {journal} {Physical Review B}\ }\textbf {\bibinfo {volume} {97}},\
  \bibinfo {pages} {174403} (\bibinfo {year} {2018})}\BibitemShut {NoStop}%
\bibitem [{\citenamefont {Long}\ \emph {et~al.}(2018)\citenamefont {Long},
  \citenamefont {Ren},\ and\ \citenamefont {Chen}}]{long2018intrinsic}%
  \BibitemOpen
  \bibfield  {author} {\bibinfo {author} {\bibfnamefont {Y.}~\bibnamefont
  {Long}}, \bibinfo {author} {\bibfnamefont {J.}~\bibnamefont {Ren}},\ and\
  \bibinfo {author} {\bibfnamefont {H.}~\bibnamefont {Chen}},\ }\bibfield
  {title} {\bibinfo {title} {Intrinsic spin of elastic waves},\ }\href@noop {}
  {\bibfield  {journal} {\bibinfo  {journal} {Proc. Natl. Acad. Sci.}\ }\textbf
  {\bibinfo {volume} {115}},\ \bibinfo {pages} {9951} (\bibinfo {year}
  {2018})}\BibitemShut {NoStop}%
\bibitem [{\citenamefont {Chaplain}\ \emph {et~al.}(2022)\citenamefont
  {Chaplain}, \citenamefont {De~Ponti},\ and\ \citenamefont
  {Craster}}]{Chaplain2022eOAM}%
  \BibitemOpen
  \bibfield  {author} {\bibinfo {author} {\bibfnamefont {G.~J.}\ \bibnamefont
  {Chaplain}}, \bibinfo {author} {\bibfnamefont {J.~M.}\ \bibnamefont
  {De~Ponti}},\ and\ \bibinfo {author} {\bibfnamefont {R.~V.}\ \bibnamefont
  {Craster}},\ }\bibfield  {title} {\bibinfo {title} {Elastic orbital angular
  momentum},\ }\href@noop {} {\bibfield  {journal} {\bibinfo  {journal} {Phys.
  Rev. Lett.}\ }\textbf {\bibinfo {volume} {128}},\ \bibinfo {pages} {064301}
  (\bibinfo {year} {2022})}\BibitemShut {NoStop}%
\bibitem [{\citenamefont {Bliokh}(2022)}]{bliokh2022elastic}%
  \BibitemOpen
  \bibfield  {author} {\bibinfo {author} {\bibfnamefont {K.~Y.}\ \bibnamefont
  {Bliokh}},\ }\bibfield  {title} {\bibinfo {title} {Elastic spin and orbital
  angular momenta},\ }\href@noop {} {\bibfield  {journal} {\bibinfo  {journal}
  {arXiv preprint arXiv:2204.13037}\ } (\bibinfo {year} {2022})}\BibitemShut
  {NoStop}%
\bibitem [{\citenamefont {Durnin}\ \emph {et~al.}(1987)\citenamefont {Durnin},
  \citenamefont {Miceli~Jr},\ and\ \citenamefont
  {Eberly}}]{durnin1987diffraction}%
  \BibitemOpen
  \bibfield  {author} {\bibinfo {author} {\bibfnamefont {J.}~\bibnamefont
  {Durnin}}, \bibinfo {author} {\bibfnamefont {J.}~\bibnamefont {Miceli~Jr}},\
  and\ \bibinfo {author} {\bibfnamefont {J.~H.}\ \bibnamefont {Eberly}},\
  }\bibfield  {title} {\bibinfo {title} {Diffraction-free beams},\ }\href@noop
  {} {\bibfield  {journal} {\bibinfo  {journal} {Phys. Rev. Lett.}\ }\textbf
  {\bibinfo {volume} {58}},\ \bibinfo {pages} {1499} (\bibinfo {year}
  {1987})}\BibitemShut {NoStop}%
\bibitem [{\citenamefont {Landau}\ and\ \citenamefont
  {Lifshitz}(1959)}]{landau1959course}%
  \BibitemOpen
  \bibfield  {author} {\bibinfo {author} {\bibfnamefont {L.~D.}\ \bibnamefont
  {Landau}}\ and\ \bibinfo {author} {\bibfnamefont {E.~M.}\ \bibnamefont
  {Lifshitz}},\ }\href@noop {} {\emph {\bibinfo {title} {Course of Theoretical
  Physics Vol 7: Theory and Elasticity}}}\ (\bibinfo  {publisher} {Pergamon
  press},\ \bibinfo {year} {1959})\BibitemShut {NoStop}%
\bibitem [{\citenamefont {Gazis}(1959{\natexlab{a}})}]{gazis1959a}%
  \BibitemOpen
  \bibfield  {author} {\bibinfo {author} {\bibfnamefont {D.~C.}\ \bibnamefont
  {Gazis}},\ }\bibfield  {title} {\bibinfo {title} {Three-dimensional
  investigation of the propagation of waves in hollow circular cylinders. {I}.
  {A}nalytical foundation},\ }\href@noop {} {\bibfield  {journal} {\bibinfo
  {journal} {J. Acoust. Soc. Am.}\ }\textbf {\bibinfo {volume} {31}},\ \bibinfo
  {pages} {568} (\bibinfo {year} {1959}{\natexlab{a}})}\BibitemShut {NoStop}%
\bibitem [{\citenamefont {Morand}\ and\ \citenamefont
  {Ohayon}(1995)}]{morand1995fluid}%
  \BibitemOpen
  \bibfield  {author} {\bibinfo {author} {\bibfnamefont {H.~J.~P.}\
  \bibnamefont {Morand}}\ and\ \bibinfo {author} {\bibfnamefont
  {R.}~\bibnamefont {Ohayon}},\ }\href@noop {} {\emph {\bibinfo {title} {Fluid
  Structure Interaction}}}\ (\bibinfo  {publisher} {John Wiley \& Sons},\
  \bibinfo {address} {Chichester},\ \bibinfo {year} {1995})\BibitemShut
  {NoStop}%
\bibitem [{\citenamefont {Berm{\'u}dez}\ \emph {et~al.}(2008)\citenamefont
  {Berm{\'u}dez}, \citenamefont {Gamallo}, \citenamefont {Hervella-Nieto},
  \citenamefont {Rodr{\'\i}guez},\ and\ \citenamefont
  {Santamarina}}]{bermudez2008fluid}%
  \BibitemOpen
  \bibfield  {author} {\bibinfo {author} {\bibfnamefont {A.}~\bibnamefont
  {Berm{\'u}dez}}, \bibinfo {author} {\bibfnamefont {P.}~\bibnamefont
  {Gamallo}}, \bibinfo {author} {\bibfnamefont {L.}~\bibnamefont
  {Hervella-Nieto}}, \bibinfo {author} {\bibfnamefont {R.}~\bibnamefont
  {Rodr{\'\i}guez}},\ and\ \bibinfo {author} {\bibfnamefont {D.}~\bibnamefont
  {Santamarina}},\ }\bibfield  {title} {\bibinfo {title} {Fluid--structure
  acoustic interaction},\ }in\ \href@noop {} {\emph {\bibinfo {booktitle}
  {Computational Acoustics of Noise Propagation in Fluids-Finite and Boundary
  Element Methods}}}\ (\bibinfo  {publisher} {Springer},\ \bibinfo {year}
  {2008})\ pp.\ \bibinfo {pages} {253--286}\BibitemShut {NoStop}%
\bibitem [{\citenamefont {COMSOL}(2021)}]{comsolSolidMech}%
  \BibitemOpen
  \bibfield  {author} {\bibinfo {author} {\bibnamefont {COMSOL}},\ }\href@noop
  {} {\emph {\bibinfo {title} {Solid Mechanics Module User's Guide}}}\
  (\bibinfo {address} {Stockholm, Sweden},\ \bibinfo {year} {2021})\BibitemShut
  {NoStop}%
\bibitem [{\citenamefont {Franke-Arnold}\ \emph {et~al.}(2008)\citenamefont
  {Franke-Arnold}, \citenamefont {Allen},\ and\ \citenamefont
  {Padgett}}]{franke2008advances}%
  \BibitemOpen
  \bibfield  {author} {\bibinfo {author} {\bibfnamefont {S.}~\bibnamefont
  {Franke-Arnold}}, \bibinfo {author} {\bibfnamefont {L.}~\bibnamefont
  {Allen}},\ and\ \bibinfo {author} {\bibfnamefont {M.}~\bibnamefont
  {Padgett}},\ }\bibfield  {title} {\bibinfo {title} {Advances in optical
  angular momentum},\ }\href@noop {} {\bibfield  {journal} {\bibinfo  {journal}
  {Laser \& Photonics Reviews}\ }\textbf {\bibinfo {volume} {2}},\ \bibinfo
  {pages} {299} (\bibinfo {year} {2008})}\BibitemShut {NoStop}%
\bibitem [{\citenamefont {Chaplain}\ and\ \citenamefont {{De
  Ponti}}(2022)}]{Chaplain2022eSPP}%
  \BibitemOpen
  \bibfield  {author} {\bibinfo {author} {\bibfnamefont {G.}~\bibnamefont
  {Chaplain}}\ and\ \bibinfo {author} {\bibfnamefont {J.}~\bibnamefont {{De
  Ponti}}},\ }\bibfield  {title} {\bibinfo {title} {The elastic spiral phase
  pipe},\ }\href {https://doi.org/https://doi.org/10.1016/j.jsv.2021.116718}
  {\bibfield  {journal} {\bibinfo  {journal} {J. Sound Vib.}\ }\textbf
  {\bibinfo {volume} {523}},\ \bibinfo {pages} {116718} (\bibinfo {year}
  {2022})}\BibitemShut {NoStop}%
\bibitem [{\citenamefont {Gazis}(1959{\natexlab{b}})}]{gazis1959b}%
  \BibitemOpen
  \bibfield  {author} {\bibinfo {author} {\bibfnamefont {D.~C.}\ \bibnamefont
  {Gazis}},\ }\bibfield  {title} {\bibinfo {title} {Three-dimensional
  investigation of the propagation of waves in hollow circular cylinders. {II}.
  {N}umerical results},\ }\href@noop {} {\bibfield  {journal} {\bibinfo
  {journal} {J. Acoust. Soc. Am.}\ }\textbf {\bibinfo {volume} {31}},\ \bibinfo
  {pages} {573} (\bibinfo {year} {1959}{\natexlab{b}})}\BibitemShut {NoStop}%
\bibitem [{\citenamefont {Silk}\ and\ \citenamefont
  {Bainton}(1979)}]{silk1979propagation}%
  \BibitemOpen
  \bibfield  {author} {\bibinfo {author} {\bibfnamefont {M.}~\bibnamefont
  {Silk}}\ and\ \bibinfo {author} {\bibfnamefont {K.}~\bibnamefont {Bainton}},\
  }\bibfield  {title} {\bibinfo {title} {The propagation in metal tubing of
  ultrasonic wave modes equivalent to {L}amb waves},\ }\href@noop {} {\bibfield
   {journal} {\bibinfo  {journal} {Ultrasonics}\ }\textbf {\bibinfo {volume}
  {17}},\ \bibinfo {pages} {11} (\bibinfo {year} {1979})}\BibitemShut {NoStop}%
\bibitem [{\citenamefont {Alleyne}\ \emph {et~al.}(1998)\citenamefont
  {Alleyne}, \citenamefont {Lowe},\ and\ \citenamefont {Cawley}}]{Alleyne98}%
  \BibitemOpen
  \bibfield  {author} {\bibinfo {author} {\bibfnamefont {D.~N.}\ \bibnamefont
  {Alleyne}}, \bibinfo {author} {\bibfnamefont {M.~J.~S.}\ \bibnamefont
  {Lowe}},\ and\ \bibinfo {author} {\bibfnamefont {P.}~\bibnamefont {Cawley}},\
  }\bibfield  {title} {\bibinfo {title} {{The Reflection of Guided Waves From
  Circumferential Notches in Pipes}},\ }\href
  {https://doi.org/10.1115/1.2789105} {\bibfield  {journal} {\bibinfo
  {journal} {J. Appl. Mech.}\ }\textbf {\bibinfo {volume} {65}},\ \bibinfo
  {pages} {635} (\bibinfo {year} {1998})}\BibitemShut {NoStop}%
\bibitem [{\citenamefont {Lowe}\ \emph {et~al.}(1998)\citenamefont {Lowe},
  \citenamefont {Alleyne},\ and\ \citenamefont {Cawley}}]{Lowe98}%
  \BibitemOpen
  \bibfield  {author} {\bibinfo {author} {\bibfnamefont {M.~J.~S.}\
  \bibnamefont {Lowe}}, \bibinfo {author} {\bibfnamefont {D.~N.}\ \bibnamefont
  {Alleyne}},\ and\ \bibinfo {author} {\bibfnamefont {P.}~\bibnamefont
  {Cawley}},\ }\bibfield  {title} {\bibinfo {title} {{The Mode Conversion of a
  Guided Wave by a Part-Circumferential Notch in a Pipe}},\ }\href
  {https://doi.org/10.1115/1.2789107} {\bibfield  {journal} {\bibinfo
  {journal} {J. Appl. Mech.}\ }\textbf {\bibinfo {volume} {65}},\ \bibinfo
  {pages} {649} (\bibinfo {year} {1998})}\BibitemShut {NoStop}%
\bibitem [{\citenamefont {Shin}\ and\ \citenamefont
  {Rose}(1999)}]{shin1999guided}%
  \BibitemOpen
  \bibfield  {author} {\bibinfo {author} {\bibfnamefont {H.~J.}\ \bibnamefont
  {Shin}}\ and\ \bibinfo {author} {\bibfnamefont {J.~L.}\ \bibnamefont
  {Rose}},\ }\bibfield  {title} {\bibinfo {title} {Guided waves by axisymmetric
  and non-axisymmetric surface loading on hollow cylinders},\ }\href@noop {}
  {\bibfield  {journal} {\bibinfo  {journal} {Ultrasonics}\ }\textbf {\bibinfo
  {volume} {37}},\ \bibinfo {pages} {355} (\bibinfo {year} {1999})}\BibitemShut
  {NoStop}%
\bibitem [{\citenamefont {Tang}\ and\ \citenamefont
  {Wu}(2017)}]{tang2017excitation}%
  \BibitemOpen
  \bibfield  {author} {\bibinfo {author} {\bibfnamefont {L.}~\bibnamefont
  {Tang}}\ and\ \bibinfo {author} {\bibfnamefont {B.}~\bibnamefont {Wu}},\
  }\bibfield  {title} {\bibinfo {title} {Excitation mechanism of
  flexural-guided wave modes {F}(1, 2) and {F}(1, 3) in pipes},\ }\href@noop {}
  {\bibfield  {journal} {\bibinfo  {journal} {J. Nondestruct. Eval.}\ }\textbf
  {\bibinfo {volume} {36}},\ \bibinfo {pages} {1} (\bibinfo {year}
  {2017})}\BibitemShut {NoStop}%
\bibitem [{\citenamefont {Lowe}(2001)}]{LOWE20011551}%
  \BibitemOpen
  \bibfield  {author} {\bibinfo {author} {\bibfnamefont {M.}~\bibnamefont
  {Lowe}},\ }\bibfield  {title} {\bibinfo {title} {Wave propagation | guided
  waves in structures},\ }in\ \href@noop {} {\emph {\bibinfo {booktitle}
  {Encyclopedia of Vibration}}},\ \bibinfo {editor} {edited by\ \bibinfo
  {editor} {\bibfnamefont {S.}~\bibnamefont {Braun}}}\ (\bibinfo  {publisher}
  {Elsevier},\ \bibinfo {address} {Oxford},\ \bibinfo {year} {2001})\ pp.\
  \bibinfo {pages} {1551--1559}\BibitemShut {NoStop}%
\bibitem [{\citenamefont {Kwun}\ \emph {et~al.}(2008)\citenamefont {Kwun},
  \citenamefont {Kim}, \citenamefont {Matsumoto},\ and\ \citenamefont
  {Vinogradov}}]{kwun2008detection}%
  \BibitemOpen
  \bibfield  {author} {\bibinfo {author} {\bibfnamefont {H.}~\bibnamefont
  {Kwun}}, \bibinfo {author} {\bibfnamefont {S.~Y.}\ \bibnamefont {Kim}},
  \bibinfo {author} {\bibfnamefont {H.}~\bibnamefont {Matsumoto}},\ and\
  \bibinfo {author} {\bibfnamefont {S.}~\bibnamefont {Vinogradov}},\ }\bibfield
   {title} {\bibinfo {title} {Detection of axial cracks in tube and pipe using
  torsional guided waves},\ }in\ \href@noop {} {\emph {\bibinfo {booktitle}
  {AIP Conference Proceedings}}},\ Vol.\ \bibinfo {volume} {975}\ (\bibinfo
  {organization} {American Institute of Physics},\ \bibinfo {year} {2008})\
  pp.\ \bibinfo {pages} {193--199}\BibitemShut {NoStop}%
\bibitem [{\citenamefont {Ratassepp}\ \emph {et~al.}(2010)\citenamefont
  {Ratassepp}, \citenamefont {Fletcher},\ and\ \citenamefont
  {Lowe}}]{ratassepp2010scattering}%
  \BibitemOpen
  \bibfield  {author} {\bibinfo {author} {\bibfnamefont {M.}~\bibnamefont
  {Ratassepp}}, \bibinfo {author} {\bibfnamefont {S.}~\bibnamefont
  {Fletcher}},\ and\ \bibinfo {author} {\bibfnamefont {M.}~\bibnamefont
  {Lowe}},\ }\bibfield  {title} {\bibinfo {title} {Scattering of the
  fundamental torsional mode at an axial crack in a pipe},\ }\href@noop {}
  {\bibfield  {journal} {\bibinfo  {journal} {J. Acoust. Soc. Am.}\ }\textbf
  {\bibinfo {volume} {127}},\ \bibinfo {pages} {730} (\bibinfo {year}
  {2010})}\BibitemShut {NoStop}%
\bibitem [{\citenamefont {Beijersbergen}\ \emph {et~al.}(1994)\citenamefont
  {Beijersbergen}, \citenamefont {Coerwinkel}, \citenamefont {Kristensen},\
  and\ \citenamefont {Woerdman}}]{beijersbergen1994helical}%
  \BibitemOpen
  \bibfield  {author} {\bibinfo {author} {\bibfnamefont {M.}~\bibnamefont
  {Beijersbergen}}, \bibinfo {author} {\bibfnamefont {R.}~\bibnamefont
  {Coerwinkel}}, \bibinfo {author} {\bibfnamefont {M.}~\bibnamefont
  {Kristensen}},\ and\ \bibinfo {author} {\bibfnamefont {J.}~\bibnamefont
  {Woerdman}},\ }\bibfield  {title} {\bibinfo {title} {Helical-wavefront laser
  beams produced with a spiral phaseplate},\ }\href@noop {} {\bibfield
  {journal} {\bibinfo  {journal} {Opt. Commum.}\ }\textbf {\bibinfo {volume}
  {112}},\ \bibinfo {pages} {321} (\bibinfo {year} {1994})}\BibitemShut
  {NoStop}%
\bibitem [{\citenamefont {Adamou}\ and\ \citenamefont
  {Craster}(2004)}]{adamou2004spectral}%
  \BibitemOpen
  \bibfield  {author} {\bibinfo {author} {\bibfnamefont {A.}~\bibnamefont
  {Adamou}}\ and\ \bibinfo {author} {\bibfnamefont {R.}~\bibnamefont
  {Craster}},\ }\bibfield  {title} {\bibinfo {title} {Spectral methods for
  modelling guided waves in elastic media},\ }\href@noop {} {\bibfield
  {journal} {\bibinfo  {journal} {J. Acoust. Soc. Am.}\ }\textbf {\bibinfo
  {volume} {116}},\ \bibinfo {pages} {1524} (\bibinfo {year}
  {2004})}\BibitemShut {NoStop}%
\bibitem [{\citenamefont {Ricci}\ \emph {et~al.}(2012)\citenamefont {Ricci},
  \citenamefont {L{\"o}ffler},\ and\ \citenamefont
  {Van~Exter}}]{ricci2012instability}%
  \BibitemOpen
  \bibfield  {author} {\bibinfo {author} {\bibfnamefont {F.}~\bibnamefont
  {Ricci}}, \bibinfo {author} {\bibfnamefont {W.}~\bibnamefont {L{\"o}ffler}},\
  and\ \bibinfo {author} {\bibfnamefont {M.}~\bibnamefont {Van~Exter}},\
  }\bibfield  {title} {\bibinfo {title} {Instability of higher-order optical
  vortices analyzed with a multi-pinhole interferometer},\ }\href@noop {}
  {\bibfield  {journal} {\bibinfo  {journal} {Opt. Express}\ }\textbf {\bibinfo
  {volume} {20}},\ \bibinfo {pages} {22961} (\bibinfo {year}
  {2012})}\BibitemShut {NoStop}%
\bibitem [{\citenamefont {Berry}\ and\ \citenamefont
  {Dennis}(2001)}]{berry2001knotted}%
  \BibitemOpen
  \bibfield  {author} {\bibinfo {author} {\bibfnamefont {M.~V.}\ \bibnamefont
  {Berry}}\ and\ \bibinfo {author} {\bibfnamefont {M.~R.}\ \bibnamefont
  {Dennis}},\ }\bibfield  {title} {\bibinfo {title} {Knotted and linked phase
  singularities in monochromatic waves},\ }\href@noop {} {\bibfield  {journal}
  {\bibinfo  {journal} {Proc. R. Soc. A}\ }\textbf {\bibinfo {volume} {457}},\
  \bibinfo {pages} {2251} (\bibinfo {year} {2001})}\BibitemShut {NoStop}%
\bibitem [{\citenamefont {Schmid}(2010)}]{schmid_2010}%
  \BibitemOpen
  \bibfield  {author} {\bibinfo {author} {\bibfnamefont {P.~J.}\ \bibnamefont
  {Schmid}},\ }\bibfield  {title} {\bibinfo {title} {Dynamic mode decomposition
  of numerical and experimental data},\ }\href
  {https://doi.org/10.1017/S0022112010001217} {\bibfield  {journal} {\bibinfo
  {journal} {J. Fluid Mech.}\ }\textbf {\bibinfo {volume} {656}},\ \bibinfo
  {pages} {5–28} (\bibinfo {year} {2010})}\BibitemShut {NoStop}%
\bibitem [{\citenamefont {Bruus}(2012)}]{bruus2012acoustofluidics}%
  \BibitemOpen
  \bibfield  {author} {\bibinfo {author} {\bibfnamefont {H.}~\bibnamefont
  {Bruus}},\ }\bibfield  {title} {\bibinfo {title} {Acoustofluidics 7: The
  acoustic radiation force on small particles},\ }\href@noop {} {\bibfield
  {journal} {\bibinfo  {journal} {Lab on a Chip}\ }\textbf {\bibinfo {volume}
  {12}},\ \bibinfo {pages} {1014} (\bibinfo {year} {2012})}\BibitemShut
  {NoStop}%
\bibitem [{\citenamefont {Toftul}\ \emph {et~al.}(2019)\citenamefont {Toftul},
  \citenamefont {Bliokh}, \citenamefont {Petrov},\ and\ \citenamefont
  {Nori}}]{toftul2019acoustic}%
  \BibitemOpen
  \bibfield  {author} {\bibinfo {author} {\bibfnamefont {I.}~\bibnamefont
  {Toftul}}, \bibinfo {author} {\bibfnamefont {K.}~\bibnamefont {Bliokh}},
  \bibinfo {author} {\bibfnamefont {M.~I.}\ \bibnamefont {Petrov}},\ and\
  \bibinfo {author} {\bibfnamefont {F.}~\bibnamefont {Nori}},\ }\bibfield
  {title} {\bibinfo {title} {Acoustic radiation force and torque on small
  particles as measures of the canonical momentum and spin densities},\
  }\href@noop {} {\bibfield  {journal} {\bibinfo  {journal} {Phys. Rev. Lett.}\
  }\textbf {\bibinfo {volume} {123}},\ \bibinfo {pages} {183901} (\bibinfo
  {year} {2019})}\BibitemShut {NoStop}%
\end{thebibliography}

%

\end{document}